\documentclass[varenna]{cimento}

\setcopyright{Lee C. Bassett}
\usepackage[sort&compress,numbers]{natbib}
\usepackage[english]{babel}

\usepackage{amsmath} 
\usepackage{amssymb}
\usepackage{amsfonts}
\usepackage{bm}
\usepackage{graphicx}  
\usepackage[cdot,mediumqspace,amssymb]{SIunits} 
\usepackage[autolanguage]{numprint} 
\usepackage[capitalise,nameinlink]{cleveref} 

\providecommand{\abs}[1]{\lvert#1\rvert}%
\providecommand{\Abs}[1]{\left\lvert#1\right\rvert}%
\newcommand\eqnref[1]{Eq.~\eqref{#1}}
\newcommand{\ket}[1]{{\left| #1 \right>}} 
\newcommand{\bra}[1]{{\left< #1 \right|}} 
\newcommand{\braket}[2]{{\left< #1 \vphantom{#2} \right| \left. #2 \vphantom{#1} \right>}} 
\newcommand{\ketbra}[2]{{\left| #1 \vphantom{#2} \right>\!\!\left< #2 \vphantom{#1} \right|}} 

\newenvironment{sistema}{%
    \left\lbrace\begin{array}{@{}l@{}}}
{\end{array}\right.}

\def\muB{\mu_\mathrm{B}}
\def\gespar{{g_\mathrm{es}^\parallel}}
\def\gesperp{{g_\mathrm{es}^\perp}}
\def\Eph{E_\mathrm{ph}}
\def\nD{n_\mathrm{D}}
\def\Aeff{A_\mathrm{eff}}
\def\Efield{\vec{E}_0}

\def\Sose{\Sigma_\mathrm{S}}
\def\PhiF{\Phi_\mathrm{F}}
\DeclareMathOperator{\Tr}{Tr}%

\title{Quantum optics with single spins}
\author{Lee C. Bassett}
\institute{Quantum Engineering Laboratory\\ Department of Electrical \& Systems Engineering\\ University of Pennsylvania\\  Philadelphia, PA, USA}

\begin{document}

\begin{center}
\textsc{Note: This is the preprint of an article to appear in the}\\[0.5\baselineskip]
\textbf{\large Proceedings of the International School of Physics ``E. Fermi''}\\
\textbf{\large Course 204: \textit{Nanoscale Quantum Optics}}\\[0.5\baselineskip]
\textsl{\large eds. M. Agio, I. D'Amico, C. Toninelli, and R. Zia}\\[0.5\baselineskip]
\end{center}

{\let\newpage\relax\maketitle}

\vspace{-10em}
\noindent\textsc{A course presented at the International School of Physics ``Enrico Fermi'' at Villa Monastero, Varenna, Lake Como, Italy in July 2018.}
\vspace{5em}

\begin{abstract}
Defects in solids are in many ways analogous to trapped atoms or molecules.
They can serve as long-lived quantum memories and efficient light-matter interfaces.
As such, they are leading building blocks for long-distance quantum networks and distributed quantum computers.
This chapter describes the quantum-mechanical coupling between atom-like spin states and light, using the diamond nitrogen-vacancy (NV) center as a paradigm.
We present an overview of the NV center's electronic structure, derive a general picture of coherent light-matter interactions, and describe several methods that can be used to achieve all-optical initialization, quantum-coherent control, and readout of solid-state spins.
These techniques can be readily generalized to other defect systems, and they serve as the basis for advanced protocols at the heart of many emerging quantum technologies.
\end{abstract}

\section{Introduction}

Solid-state spins are among the most versatile platforms for quantum science and technology. 
Select semiconductor defects --- exemplified by the nitrogen-vacancy (NV) center in diamond --- exhibit spin coherence at room-temperature and intrinsic optical spin-readout mechanisms that underly their remarkable capabilities as room-temperature qubits and quantum sensors. 
When used in this way, quantum-coherent control is performed using microwaves that couple resonantly to the qubit's electron spin Hamiltonian.
Optical pumping and fluorescence are used for spin initialization and readout, respectively, but these processes rely on dissipation through nonradiative and vibronic transitions that involve coupling to phonons in the crystal and are therefore incoherent.
When the crystal is cooled down, however, the optical transitions between different orbital states become coherent, and they can be manipulated using resonant optical fields just as the spin is controlled with microwaves.
Moreover, spin-orbit coupling mediates interactions between optical fields and spins, enabling all-optical (i.e., microwave free) spin control, robust spin initialization and readout, and various schemes for generating spin-photon entanglement.

In this chapter based on lectures from the 2018 Enrico Fermi Summer School on \emph{Nanoscale Quantum Optics}, we introduce a general picture for coherent light-matter interactions based on coherent, dispersive interactions with spin-selective optical transitions based on the Jaynes-Cummings Hamiltonian for quantum electrodynamics.
The chapter is organized as follows.
In \cref{sec:NVcenter}, we introduce the spin and optical fine structure of the NV center, including the role of various perturbations.
Subsequently, in \cref{sec:LightMatterCoupling} we summarize key concepts of quantum optics including the Jaynes-Cummings Hamiltonian, and show how coherent, dispersive, light-matter interactions give rise to the optical Stark effect and the Faraday effect, which can be used respectively to control and measure NV-center spin states.
In \cref{sec:CPT-SRT} we generalize the treatment to include more complex dynamics exhibited by an optical $\Lambda$ configuration, including coherent population trapping and stimulated Raman transitions, and in \cref{sec:Ultrafast} we describe an alternate, non-dispersive technique to probe and control quantum dynamics using ultrafast optical pulses.
\Cref{sec:Summary} summarizes the chapter and highlights future directions for the application of these techniques to address other spin-qubit platforms, and to enable more advanced schemes for quantum control within quantum networks.
Much of the material is adapted from Buckley \textit{et al.} \cite{Buckley2010}, Yale \textit{et al.} \cite{Yale2013}, and Bassett \textit{et al.} \cite{Bassett2014}, and more information regarding the experiments and models can be found in those references.

\section{Electronic structure of the diamond nitrogen-vacancy center\label{sec:NVcenter}}

\begin{figure}
\begin{center}
\includegraphics[width=\textwidth]{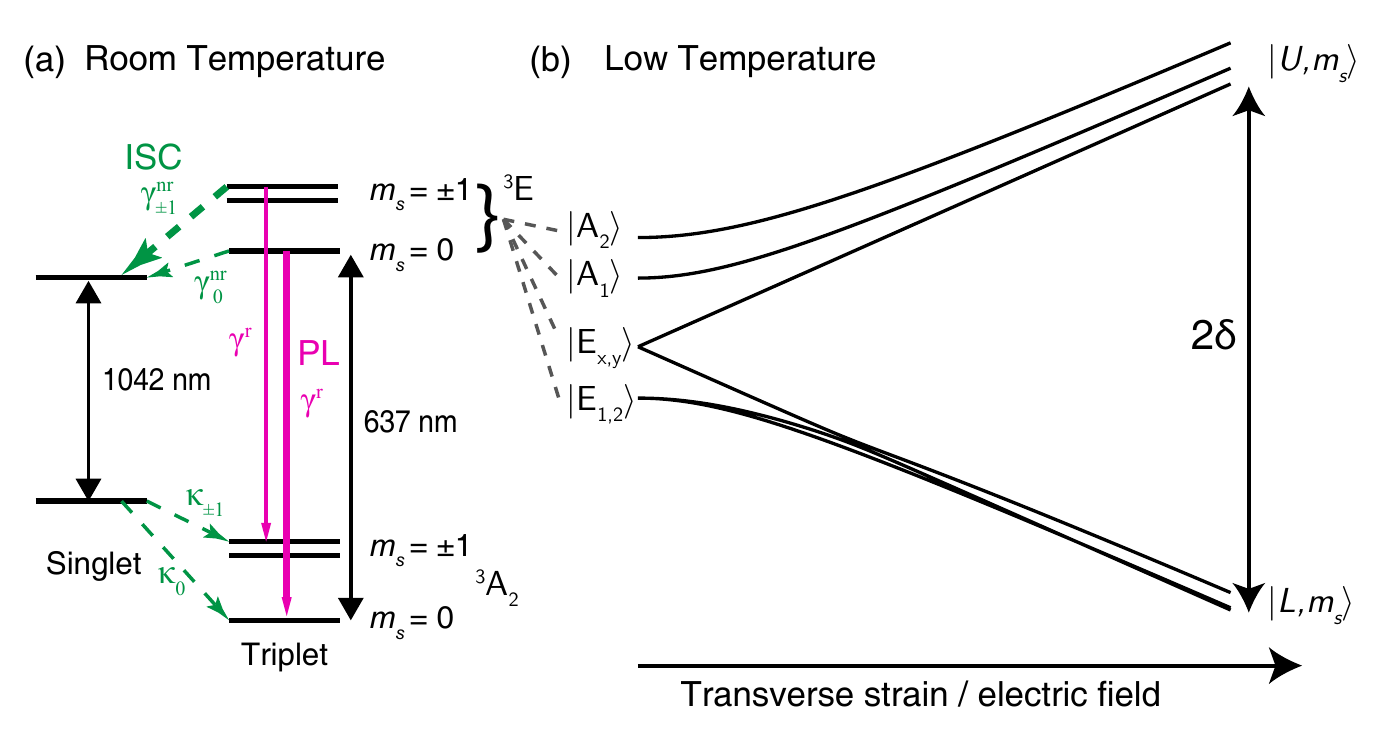}
  \caption[Electronic structure of the NV center]{\label{fig:NVstructure}
  \textbf{Electronic structure of the diamond NV center}. (a) At room temperature, rapid phonon transitions within the orbitals of the $^{3}E$ excited state lead to an effective spin-triplet with spin-dependent non-radiate decay channels, $\gamma^\mathrm{nr}_{m_s}$, through the ISC as shown. These dynamics produce the NV center's spin-dependent PL. (b) At low temperature, the orbital fine structure within $^{3}E$ is resolved. The unperturbed spin-orbit states evolve into two separated orbital branches as a function of the transverse strain or dc Stark shift, $\delta$. Eigenstates in (b) are calculated according to \cref{eq:Hes} for $B=0$~G and $\alpha_s=0$.}
\end{center}
\end{figure}

The NV center in diamond has been an object of fascination since the 1950s as one of the predominant color centers in diamond, and the focus of intense study in quantum information science since the turn of the 21$^\mathrm{st}$ century \cite{Doherty2013}.
Its popularity and importance in quantum science stem from several key characteristics, including long spin coherence of the triplet ground state, which persists to room temperature and above, and efficient, stable, visible photoluminescence (PL) that can be used to measure the spin state populations.
The latter property stems from the NV center's specific electronic level structure, which at room temperature takes the effective form shown in \cref{fig:NVstructure}(a).
The spin-triplet ground state and optically excited state\,---\,which is responsible for the visible PL\,---\,is connected to manifold of intermediate singlet states through an inter-system crossing (ISC).
The nonradiative ISC is mediated by phonons and the spin-orbit interaction, and the rates in both the upper and lower branches depend on the triplet spin projection.
In particular, the upper ISC transition from the triplet excited state occurs predominantly for the $m_s=\pm1$ spin sublevels (labeled according to the $S_z$ projection, where $z$ is along the defect's symmetry axis).
These intrinsic, spin-dependent optical dynamics provide the mechanisms for optical spin initialization and PL-based spin readout that are used in a majority of NV-center applications, especially at room temperature \cite{Hopper2018_spinreadout}.

While the $^{3}A_2$ ground state is an orbital singlet, the $^{3}E$ excited state is an orbital doublet.
At room temperature, rapid phonon-mediated transitions between the orbital branches result in an effective spin-triplet Hamiltonian as shown in \cref{fig:NVstructure}(a).
At temperatures below $\approx$\unit{20}{\kelvin}, however, phonon-induced transitions are suppressed, and the fine structure associated with the full six-dimensional excited-state Hamiltonian emerges in the optical transitions, as shown in \cref{fig:NVstructure}(b) \cite{Fu2009,Batalov2009}.
At these temperatures, and in pure diamond samples, the zero-phonon-line (ZPL) transitions become spectrally narrow, in some cases approaching the lifetime limit, such that coherent Rabi oscillations can be observed between the ground and excited-state orbitals \cite{Robledo2010}.

\subsection{The electronic Hamiltonian}\label{sec:NVHamiltonian}

The form of the NV center's electronic Hamiltonian can be derived and understood using group theory \cite{Maze2011,Doherty2011}.
The ground-state Hamiltonian is given by
\begin{equation}
  H_\mathrm{gs} = D_\mathrm{gs} \left(S_z^2 -\frac{2}{3}\right) + g_\mathrm{gs}\muB \mathbf{S}\cdot\mathbf{B},
\end{equation}
where $D_\mathrm{gs}$ is the reduced matrix element (RME) for the axial spin-spin interaction, and the second term describes the Zeeman interaction in terms of the Land\'{e} $g$-factor, $g_\mathrm{gs}$, Bohr magneton $\muB$, electron spin operator $\mathbf{S}$ (where $S^2=1$, $S_\pm=S_x\pm iS_y$), and magnetic field $\mathbf{B}$.
We generally set $h=1$ such that terms in the Hamiltonian can be written in either energy or frequency units.
Similarly, the excited-state Hamiltonian can be written as a sum of terms due to intrinsic (spin-orbit, spin-spin) interactions and extrinsic (magnetic, strain/electric) fields,
\begin{equation}\label{eq:Hes}
  H_\mathrm{es} = H_\mathrm{so}+H_\mathrm{ss}+H_\mathrm{Z}+H_L+H_\mathrm{s}.
\end{equation}
Below we give explicit matrix expressions for these terms in the product basis $\ket{\varepsilon,m_s} \in \{(\ket{X},\ket{Y})\otimes(\ket{-1},\ket{0},\ket{+1})\}$, where $(\ket{X},\ket{Y})$ are $E$-symmetry orbital states transforming like the
vectors $(x,y)$ in the NV-center coordinate system.
The spin sublevels $\ket{m_s}$ are eigenstates of the $S_z$ operator, whereas the orbital part of the Hamiltonian can be written in terms of the Pauli matrices $\sigma^\mathrm{es}_i$ ($i=x,y,z$) and $\sigma^\mathrm{es}_\pm=\sigma^\mathrm{es}_z\pm i\sigma^\mathrm{es}_x$, which operate on the two-dimensional orbital excited-state degree of freedom, \textit{i.e.}, $\sigma^\mathrm{es}_z|X\rangle = |X\rangle$ and $\sigma^\mathrm{es}_z|Y\rangle = -|Y\rangle$. Note that $\sigma^\mathrm{es}_\pm$ are not the standard raising and lowering operators.

The only spin-orbit coupling allowed by symmetry is the axial one (\textit{i.e.}, proportional to $S_z$ \cite{Doherty2011}), with RME $\lambda$. In the product basis, this takes the form:
\begin{equation}
  H_\mathrm{so} = -\lambda\sigma^\mathrm{es}_y S_z = \left(
  \begin{array}{cccccc}
           0 & 0 & 0 & -i\lambda & 0 & 0 \\
           0 & 0 & 0 & 0 & 0 & 0 \\
           0 & 0 & 0 & 0 & 0 & i\lambda \\
           i\lambda & 0 & 0 & 0 & 0 & 0 \\
           0 & 0 & 0 & 0 & 0 & 0 \\
           0 & 0 & -i\lambda & 0 & 0 & 0
  \end{array} \right).
\end{equation}
The spin-spin interaction has three symmetry-allowed RMEs, corresponding to one axial ($D_\mathrm{es}$) and two transverse couplings $(\Delta_1, \Delta_2)$. This takes the form:
\begin{eqnarray}\label{eq:Hss}
    H_\mathrm{ss} & = &  D_\mathrm{es} \left(S_z^2 -\frac{2}{3}\right)
    - \frac{\Delta_1 }{4}  \left( S_+^2 \sigma^\mathrm{es}_- +  S_-^2\sigma^\mathrm{es}_+\right)
    +\frac{\Delta_2}{2\sqrt{2}}
    \left(
     \left\{ S_+, S_z\right\}\sigma^\mathrm{es}_+
    +\left\{ S_-, S_z\right\}\sigma^\mathrm{es}_-
    \right) \nonumber\\
    & = & \left(
    \begin{array}{cccccc}
      D_\mathrm{es}/3 & -\Delta_2/2 & -\Delta_1/2 & 0 & i\Delta_2/2 & -i\Delta_1/2 \\
      -\Delta_2/2 & -2D_\mathrm{es}/3 & \Delta_2/2 & -i\Delta_2/2 & 0 & -i\Delta_2/2 \\
      -{\Delta_1}/{2} & {\Delta_2}/{2} & {D_\mathrm{es}}/{3} & {i\Delta_1}/{2} & {i\Delta_2}/{2} & 0 \\
      0 & {i\Delta_2}/{2} & -{i\Delta_1}/{2} & {D_\mathrm{es}}/{3} & {\Delta_2}/{2} & {\Delta_1}/{2} \\
      -{i\Delta_2}/{2} & 0 & -{i\Delta_2}/{2} & {\Delta_2}/{2} & -{2D_\mathrm{es}}/{3} & -{\Delta_2}/{2} \\
      {i\Delta_1}/{2} & {i\Delta_2}/{2} & 0 & {\Delta_1}/{2} & -{\Delta_2}/{2} & {D_\mathrm{es}}/{3}
    \end{array}\right).
\end{eqnarray}
Here, $\{A,B\}\equiv AB+BA$ denotes the anticommutator.

The Zeeman ($H_\mathrm{Z}$), diamagnetic ($H_L$) and strain/dc-Stark ($H_\mathrm{s})$ perturbations affect only the spin or orbital degrees of freedom individually.
The $E$ symmetry of the excited state allows different effective $g$-factors ($\gespar,\gesperp$) for axial and transverse components of the
Zeeman interaction, such that
\begin{eqnarray}
  H_\mathrm{Z} & = & \gesperp \muB (B_x S_x+B_y S_y) + \gespar \muB B_z S_z \nonumber\\
  & = & I_2\otimes\muB\left(
  \begin{array}{ccc}
    -\gespar B_z & \gesperp(B_x+iB_y) & 0 \\
    \gesperp(B_x-iB_y) & 0 & \gesperp(B_x+iB_y) \\
    0 & \gesperp(B_x-iB_y) & \gespar B_z
  \end{array}\right).
\end{eqnarray}
Similarly, the axial diamagnetic shift is given by the orbital operator
\begin{equation}\label{eq:HLz}
  H_L = \muB L_z B_z \sigma^\mathrm{es}_y = \muB L_z B_z\left(
  \begin{array}{cc}
    0 & -i \\
    i & 0
  \end{array}\right)\otimes I_3,
\end{equation}
where $L_z\muB$ is the $z$ component of the orbital magnetic moment.
Symmetry implies that transverse diamagnetic components are zero.
The orbital magnetic moment is known to be relatively small from measurements of circular dichroism \cite{Reddy1987,Rogers2009}, with a value $L_z=0.05$ that corresponds to a frequency shift of only $L_z \muB B/h \approx 50$~MHz at $B=100$~G.
This value is comparable to typical optical linewidths and smaller than most other terms in the Hamiltonian, hence the diamagnetic shift is often ignored for measurements performed at relatively low magnetic fields.

Finally, the perturbation due to transverse strain or electric fields is given by
\begin{equation}\label{eq:TransverseStrain}
  H_\mathrm{s} =  -\delta_x \sigma^\mathrm{es}_z  +\delta_y \sigma^\mathrm{es}_x
   =  \delta\left(
  \begin{array}{cc}
    -\cos(\alpha_s) & \sin(\alpha_s) \\
    \sin(\alpha_s) & \cos(\alpha_s)
  \end{array}\right)\otimes I_3,
\end{equation}
where $\delta_x = \delta \cos(\alpha_s)$ and $\delta_y
= \delta \sin(\alpha_s)$ are the strain (or dc Stark) perturbation components in crystallographic
$x$ and $y$ directions with units of energy, where the total transverse perturbation has an effective angle $\alpha_s$ (note the total energy splitting between the orbital branches is $2\delta$).

\subsection{Low- and high-strain regimes\label{sec:NVstrain}}
An arbitrary crystal strain tensor can be decomposed into components that transform according to the $C_{3v}$ irreducible representations $A_1$ (transforming like the vector $z$), and $E$ (with components $\{E_x,E_y\}$ that transform like the vectors $\{x,y\}$, respectively).
Similarly, the dc Stark perturbation due to electric fields applied along $z$ transform like $A_1$ whereas transverse fields transform like $E$.
Since the perturbations affect the excited-state Hamiltonian in exactly the same way, the dc Stark effect can be used to compensate an uncontrolled local strain \cite{Bassett2011}.
Whereas transverse strain/Stark fields shift the orbital energies and eigenstates according to \cref{eq:TransverseStrain}, the longitudinal perturbation is proportional to the orbital identity operator, amounting to an overall shift of the optical transition frequency between the ground and excited state, but no variations of the eigenstates within the excited-state manifold.
The longitudinal shift can be important when multiple NV centers need to be tuned to interact with indistinguishable photons \cite{Bernien2013}, however for control of individual defects it is generally possible to compensate this shift by tuning the laser, so this term is neglected here.

Near $\delta=0$, it is convenient to use the spin-orbit basis in which the Hamiltonian is nearly diagonal \cite{Doherty2011,Maze2011}, aside from the
small spin-spin coupling $\Delta_2$.
The spin-orbit basis states can be written as follows in terms of the product basis:
\begin{subequations}\label{eq:SObasis}
  \begin{align}
    & \ket{A_1}  = -\frac{i}{2}(\ket{X,-1}+\ket{X,+1}+i\ket{Y,-1}-i\ket{Y,+1}) \\
    & \ket{A_2}  = \frac{1}{2}(\ket{X,-1}-\ket{X,+1}+i\ket{Y,-1}+i\ket{Y,+1}) \\
    & \ket{E_1} \equiv \ket{E_{\pm,x}}  = -\frac{i}{2}(\ket{X,-1}+\ket{X,+1}-i\ket{Y,-1}+i\ket{Y,+1}) \\
    & \ket{E_2} \equiv \ket{E_{\pm,y}}  = -\frac{1}{2}(\ket{X,-1}-\ket{X,+1}-i\ket{Y,-1}-i\ket{Y,+1}) \\
    & \ket{E_x} \equiv \ket{E_{0,x}}  = -\ket{Y,0} \\
    & \ket{E_y} \equiv \ket{E_{0,y}}  = \ket{X,0}.
  \end{align}
\end{subequations}
In this basis, the states are labeled according to the symmetry of the tensor product of spin and orbital states, which can be obtained from tables of group-theoretic coupling coefficients \cite{Koster1963}.
For example, the state $\ket{A_1}$ transforms like the irreducible representation $A_1$.
The arrangement of these levels at zero strain and zero magnetic field is shown in \cref{fig:NVstructure}(b).
It is important to work in this low-strain regime for some applications.
For example, spin-orbit optical selection rules that link particular spin states with circular polarization states are present when $\ket{A_1}$ and $\ket{A_2}$ are excited-state eigenstates, and these selection rules can be used to generate spin-photon entanglement \cite{Togan2011} or to map photon states onto spin states \cite{Yang2016a}.

On the other hand, when the transverse strain/Stark perturbation is large, the excited-state manifold splits into two orbital branches, each with (spin-independent) linear polarization optical selection rules for transitions to the ground state.
This situation occurs when the strain splitting, $2\delta$, dominates over the other coupling terms between the orbital branches, the most important being the spin-orbit parameter $\lambda=\unit{5.33}{\giga\hertz}$ \cite{Bassett2014}.
Since strain splittings observed for NV centers in high-quality bulk diamond typically range between 5--\unit{50}{\giga\hertz}, this is often the natural situation for experiments, and it can be useful when one wishes to isolate the role of a single orbital branch.

Below, we use the Schrieffer-Wolff transformation to derive approximate expressions for the Hamiltonian in each orbital branch in this regime.
Rotating the basis in orbital space by the angle $\alpha_s$ enclosed by the crystallographic $x$ axis and the direction of the transverse perturbation, we rewrite the Hamiltonian in the form
\begin{eqnarray}
\tilde{H} &=& e^{-i\alpha_s \sigma^\mathrm{es}_y}H_\mathrm{es}e^{ i\alpha_s \sigma^\mathrm{es}_y} \nonumber\\
   &=&  g \mu_B B S_z -\lambda \sigma^\mathrm{es}_y S_z
-\delta \sigma^\mathrm{es}_z \nonumber  \\
& & +D_{\rm es} \left(S_z^2 -\frac{2}{3}\right)
- \frac{\Delta_1 }{4}  \left( e^{-i\alpha_s} S_+^2 \sigma^\mathrm{es}_- + e^{i\alpha_s} S_-^2\sigma^\mathrm{es}_+\right) \nonumber\\
& & +\frac{\Delta_2}{2\sqrt{2}}
\left(
 e^{i\alpha_s} \left\{ S_+, S_z\right\}\sigma^\mathrm{es}_+
+e^{-i\alpha_s} \left\{ S_-, S_z\right\}\sigma^\mathrm{es}_-
\right)  ,
\end{eqnarray}
where the strain term is block diagonal.
Note that in this expression we have assumed that the magnetic field is applied along $z$, and we ignore the orbital diagmagnetic shift.
In this basis, the states are labeled $\ket{\varepsilon,m_s}$, where $\varepsilon\in\{L,U\}$ are the lower-energy and higher-energy states, respectively, and $m_s\in\{-1,0,+1\}$.

Provided that $2\delta>\lambda$, we can treat the inter-branch coupling as a perturbation, dividing the Hamiltonian into
\begin{equation}
\tilde{H}= H_0+V,
\end{equation}
with the inter-branch coupling term
\begin{equation}
V = \left[-\lambda  S_z
+ \frac{i\Delta_1 }{4}  \left( e^{-i\alpha_s} S_+^2 - e^{i\alpha_s} S_-^2\right)
+\frac{i\Delta_2}{2\sqrt{2}}
\left(
 e^{i\alpha_s} \left\{ S_+, S_z\right\}
-e^{-i\alpha_s} \left\{ S_-, S_z\right\}
\right)\right]  \sigma^\mathrm{es}_y .
\end{equation}
Starting from this model, we apply quasi-degenerate perturbation theory in the form of a Schrieffer-Wolff transformation
\begin{eqnarray}
H_{\rm eff} &=& e^G \tilde{H} e^{-G} = \tilde{H} + \left[G,\tilde{H}\right]+
\frac{1}{2}\left[G,\left[G,\tilde{H}\right]\right]+O(G^3)\nonumber\\
&=&  H_0 + V + \left[G,H_0\right]+ \left[G,V\right]
+\frac{1}{2}\left[G,\left[G,H_0\right]\right] + \cdots,
\end{eqnarray}
where the generator $G$ is defined such that $G^\dagger=-G$ in order to eliminate the couplings between the two strain-split branches in lowest order.
The condition for this to work is $[G,H_0]=-V$, because it implies a transformed effective Hamiltonian
\begin{equation}
H_{\rm eff} =H_0 + \frac{1}{2}\left[G,V\right]
\end{equation}
which is second order in the couplings $\lambda$, $\Delta_1$, and $\Delta_2$.
This effective Hamiltonian is block-diagonal, \textit{i.e.}, it can be split up into a lower and upper branch component, each containing the contributions due to virtual transitions \textit{via} the other branch up to linear order in the couplings.

The effective Hamiltonian takes the general form:
\begin{equation}
H_{\rm eff} = D_{\rm es} \left(S_z^2
  -\frac{2}{3}\right)  +g \mu_B B S_z - \delta \sigma^\mathrm{es}_z +  \left(\begin{array}{c c} H_L & 0\\ 0 & H_U\end{array}\right).
\end{equation}
Within the lengthy expressions for $H_L$ and $H_U$, we assume the strain splitting $2\delta$ is the dominant energy scale, and expand to lowest order in $1/\delta$ to obtain
\begin{equation}
H_L \simeq \left(
\begin{array}{ccc}
- \frac{\lambda ^2}{2\delta} -\frac{\Delta_1^2}{8\delta}
& -\frac{1}{8} e^{-2 i \alpha_s } \Delta_2 f_+(\alpha_s)
&   \frac{1}{2} e^{-i \alpha_s } \Delta_1 \left(\frac{\lambda }{\delta }-1\right) \\
 -\frac{1}{8} e^{2 i \alpha_s } \Delta_2 f_+(\alpha_s)^*
& 0
& \frac{1}{8} e^{-2 i \alpha_s } \Delta_2 f_+(\alpha_s) \\
 \frac{1}{2} e^{i \alpha_s} \Delta_1 \left(\frac{\lambda
   }{\delta }-1\right)
& \frac{1}{8} e^{2i \alpha_s }
   \Delta_2 f_+(\alpha_s)^*
&
 - \frac{\lambda ^2}{2 \delta }-\frac{\Delta_1^2}{8 \delta }
\end{array}
\right)
+O\left(\frac{1}{\delta^2}\right),
\label{eq:HL}
\end{equation}
for the lower branch and
\begin{equation}
H_U \simeq \left(
\begin{array}{ccc}
\frac{\lambda ^2}{2\delta}  +\frac{\Delta_1^2}{8 \delta }
&
   \frac{1}{8} e^{-2 i \alpha_s } \Delta_2 f_-(\alpha_s)
& \frac{1}{2} e^{-i \alpha_s } \Delta_1
   \left(\frac{\lambda }{\delta }+1\right) \\
 \frac{1}{8} e^{2i \alpha_s } \Delta_2 f_-(\alpha_s)^*
& 0
& -\frac{1}{8} e^{-2 i \alpha_s } \Delta_2 f_-(\alpha_s) \\
 \frac{1}{2} e^{i \alpha_s } \Delta_1 \left(\frac{\lambda }{\delta}+1\right)
& -\frac{1}{8} e^{2i \alpha_s }  \Delta_2 f_-(\alpha_s)^*
&
   \frac{\lambda ^2}{2 \delta } +\frac{\Delta_1^2}{8 \delta }
\label{eq:HU}
\end{array}
\right)
+O\left(\frac{1}{\delta^2}\right)
\end{equation}
for the upper branch.
Here, we have introduced the expression $f_\pm(\alpha_s)=
\frac{\Delta_1}{\delta }+2 e^{3 i \alpha_s } \left(2\pm\frac{\lambda
       }{\delta }\right)$,
which leads to an oscillation with the strain angle $\alpha_s$ of the splitting between the $S_z=0$ and $S_z=\pm 1$ states at their respective crossing points.
The diagonal elements in Eqs.~(\ref{eq:HL}) and (\ref{eq:HU}) are the spin-orbit and spin-spin induced level repulsions between the two branches, while the off-diagonal elements are second-order spin-flip terms.

\section{Coherent light-matter coupling\label{sec:LightMatterCoupling}}

Experiments that probe spin-light coherence \cite{Buckley2010,Togan2010}, and related protocols for all-optical coherent control \cite{Yale2013,Bassett2014} of NV-center spins draw on a rich history in quantum optics and atomic physics.
For general background in this subject, we refer the reader to excellent textbooks such as those by Gerry and Knight \cite{Gerry2005} or Cohen-Tannoudji and Gu\'ery-Odelin \cite{Cohen-Tannoudji2011}.
In this section, we give a brief introduction to the concept of coherent coupling between a light field and atomic transitions, using the Jaynes-Cummings Hamiltonian to derive expressions for the optical (ac) Stark effect and the Faraday effect.
This derivation naturally captures the correspondence between these two effects, which both result from the polariton energy shifts due to the interactions between the light field and atomic transitions.
We discuss how the concept was applied by Buckley \textit{et al.} \cite{Buckley2010} to demonstrate dispersive optical measurements of the spin state and all-optical coherent spin rotations.

\subsection{The Jaynes-Cummings Hamiltonian}

The Jaynes-Cummings Hamiltonian describes the interaction between light and matter in the rotating wave approximation (see, \textit{e.g.}, Chapter 4 of Gerry and Knight \cite{Gerry2005} for a full derivation).
It is typically used in the context of cavity quantum electrodynamics to describe coherent coupling of an atom-like system to the optical field in a cavity, however it can be applied more generally even when cavities are not involved.
For example, in the experiments by Buckley \textit{et al.} \cite{Buckley2010} and Yale \textit{et al.} \cite{Yale2013}, the `cavity' is defined by the duration of a laser pulse, $\tau$, which is assumed to propagate in a single spatial mode that we can treat as a coherent state of light with a well-defined electric-field amplitude and phase.
We assume that the turn-on and turn-off of this pulse is smooth, such that the interaction with the NV center is adiabatic.
We also neglect spontaneous emission and other forms of decoherence such as spectral hopping and laser noise.
A treatment of these effects can be found in Ref.~\cite{Buckley2010}.

Our starting point is the dipole interaction Hamiltonian
\begin{equation}
  H_\mathrm{int}= \sqrt{F_\mathrm{DW}}\vec{\mu}\cdot\vec{E},
\end{equation}
where $\vec{\mu}$ is the NV-center electric dipole, $\vec{E}$ is the local electric field, and $F_\mathrm{DW}\!=\!0.04\pm0.01$ is the Debye-Waller factor, which empirically accounts for the reduced resonant coupling between NV center ground and excited states due to displacement of the nuclear coordinates during optical transitions \cite{Davies1974}.
The dipole magnitude is directly related to the NV center's spontaneous decay rate $\gamma^{r}=1/13$~\nano\reciprocal\second\ \cite{Collins1983} through
\begin{equation}
  \left|\vec{\mu}\right|^2 = \frac{3\pi\varepsilon_0\hbar^4c^3\gamma^r}{\Eph^3\nD},
\end{equation}
where $\Eph = 1.945$~\electronvolt\ is the photon energy and $\nD=2.4$ is the refractive index of diamond.
The amplitude of the electric field can be expressed in terms of the total number of photons $n$ in the pulse and the effective equal-intensity optical mode area at the NV center $\Aeff$ through the classical irradiance
\begin{equation}
  I = \frac{c\nD\varepsilon_0}{2}\big|\Efield\big|^2 = \frac{n\Eph}{\tau \Aeff},
\end{equation}
such that
\begin{equation}
  \big|\Efield\big| = \sqrt{\frac{2n\Eph}{\nD\varepsilon_0\Aeff c\tau}}.
\end{equation}
By introducing the operators $\hat{\vec{E}} = i\big|\Efield\big|(\hat{a}^\dag-\hat{a})$ and $\hat{\vec{\mu}} = \left|\vec{\mu}\right|(\hat{\sigma}_+ +\hat{\sigma}_-)$ for the electric field and optical dipole, respectively, we cast $H_\mathrm{int}$ into the form
\begin{equation}\label{eq:Hint}
  \hat{H}_\mathrm{int} \simeq i\frac{\hbar\Omega_0}{2}\left(\hat{a}^\dag\hat{\sigma}_- -\hat{a}\hat{\sigma}_+\right),
\end{equation}
where $\hat{a}^\dag$ ($\hat{a}$) and $\hat{\sigma}_+$ ($\hat{\sigma}_-$) are creation (annihilation) operators for optical photons and atomic excitations, respectively.
If the atomic ground and excited states are $\ket{g}$ and $\ket{e}$, then $\hat{\sigma}_+=\ket{e}\bra{g}$ and $\hat{\sigma}_-=\ket{g}\bra{e}$.
Here we neglect energy-nonconserving terms $\left\{\hat{a}\hat{\sigma}_-,\hat{a}^\dag\hat{\sigma}_+\right\}$ in the rotating wave approximation.
The quantity $\Omega_0$ is the on-resonance optical Rabi frequency, given by
\begin{equation}\label{eq:OpticalRabi}
  \Omega_0 = \frac{\sqrt{F_\mathrm{DW}}\left|\vec{\mu}\right|\big|\Efield\big|\cos(\theta)}{\hbar},
\end{equation}
where $\theta$ is the angle between the optical dipole and the light's linear polarization axis.
With the addition of the non-interacting Hamiltonian for the spin and light fields, given by
\begin{equation}
  \hat{H}_0 = \Eph\left(\hat{a}^\dag\hat{a}+\frac{1}{2}\right) + E_j\frac{\hat{\sigma}^{(j)}_z}{2},
\end{equation}
where $E_j$ is the transition energy for the spin state with $m_s=j$ and $\hat{\sigma}^{(j)}_z = \ket{e_j}\bra{e_j}-\ket{g_j}\bra{g_j}$ describes the NV center orbital excitation, we obtain the Jaynes-Cummings Hamiltonian describing the light-matter system when the spin is in state $j$,
\begin{align}
  \hat{H}_\mathrm{JC}^{(j)} & = \hat{H}^{(j)}_0+\hat{H}^{(j)}_\mathrm{int} \\
  & = \Eph\hat{a}^\dag\hat{a} + E_j\frac{\hat{\sigma}^{(j)}_z}{2}+\frac{\hbar\Omega_0}{2}\left(\hat{a}\hat{\sigma}^{(j)}_+ +\hat{a}^\dag\hat{\sigma}^{(j)}_-\right),
\end{align}
where we have set the optical zero-field energy to zero for simplicity.

The Hamiltonian $\hat{H}_\mathrm{JC}$ is naturally expressed in the basis of non-interacting polariton states,
\begin{equation}
\begin{sistema}
  |\psi_0^{(n,j)}\rangle = \ket{g_j}\otimes\ket{n+1} \\
  |\psi_1^{(n,j)}\rangle = \ket{e_j}\otimes\ket{n},
\end{sistema}
\end{equation}
where $\ket{g_j}$ ($\ket{e_j}$) are the bare ground (excited) states of the NV-center orbital transition for $m_s\!=\!j$, and $\ket{n}$ is a photon-number Fock state of the electromagnetic field.
By diagonalizing $\hat{H}_\mathrm{JC}$ in this basis, we obtain the eigenenergies
\begin{equation}
  E_{\pm}(n,\Delta_j) = \Eph\left(n+\frac{1}{2}\right)\pm\frac{\hbar}{2}\sqrt{\Delta_j^2+\Omega_0^2(n)},
\end{equation}
where $\Delta_j = (\Eph-E_j)/\hbar$ is the detuning of the laser from the unshifted NV-center transition frequency and the $n$-dependence of $\Omega_0$ (implicit through $\big|\Efield\big|$ in Eq.~\ref{eq:OpticalRabi}) is shown explicitly.
These eigenenergies take the form of an anticrossing about $\Delta_j=0$.

\begin{figure}
\begin{center}
\includegraphics[width=\textwidth]{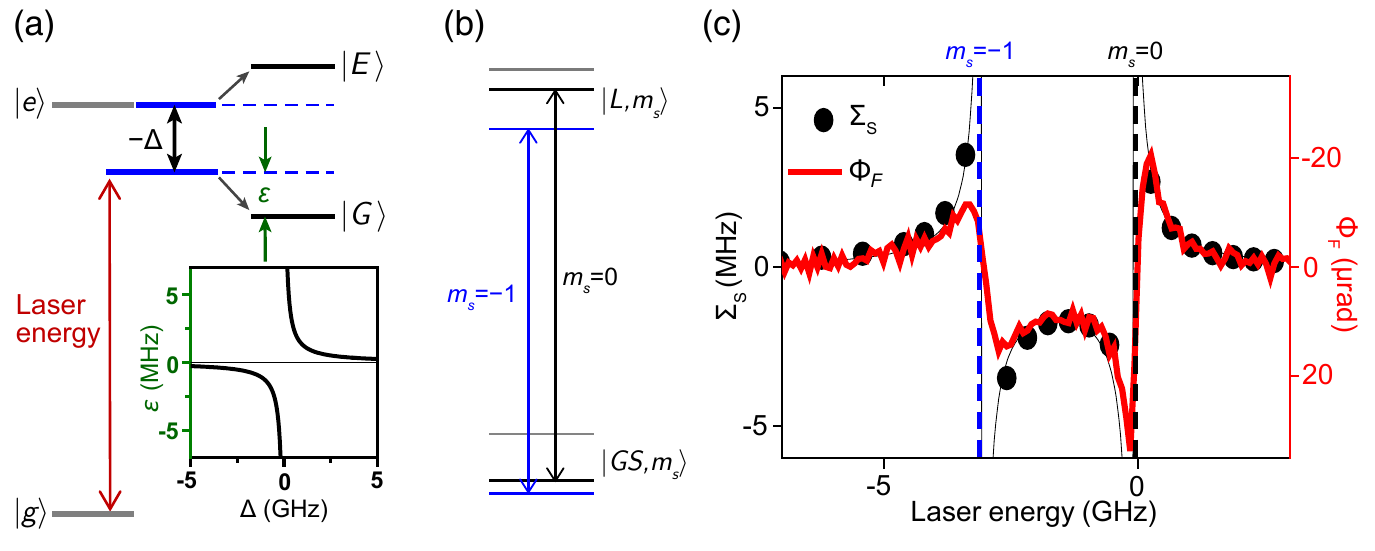}
  \caption[Light-matter coupling]{\label{fig:LightMatterCoupling}
  \textbf{Light-matter coupling in the diamond NV center}. (a) The interaction between an atomic transition with ground and excited states $\{\ket{g},\ket{e}\}$ and a near-resonant laser field is described by the Jaynes Cummings Hamiltonian in terms of polariton states $\{\ket{G},\ket{E}\}$ with an energy shift $\varepsilon$. (b) Energy-resolved transitions for different spin sublevels in the NV center's optical fine structure produce spin-dependent interactions, which manifest (c) as optical Stark rotations with frequency $\Sigma_\mathrm{S}$ and a Faraday phase shift, $\Phi_\mathrm{F}$ as a function of laser energy. Panels (a) and (c) are adapted from Ref. \cite{Buckley2010} and reprinted with permission from AAAS.}
\end{center}
\end{figure}

Since the atom is initially in the ground state and we assume that the onset of the light field is adiabatic, the occupied state during the pulse will be the polariton eigenstate having maximum overlap with $|\psi_0\rangle$, which has energy $E_g = E_\pm$ for $\Delta_j\!\gtrless0$.
The observed energy shift of this `$\ket{g_j}$-like' state relative to its non-interacting energy
\begin{equation}
  E_{g0} = \Eph(n+1)-\frac{E_j}{2}
\end{equation}
is therefore given by
\begin{equation}\label{eq:EnergyShift}
  \varepsilon_g(n,\Delta_j) = E_g-E_{g0} =  \frac{\hbar\Delta_j}{2}\left[\sqrt{1+\frac{\Omega_0^2}{\Delta_j^2}}-1\right],
\end{equation}
and is plotted in \cref{fig:LightMatterCoupling}(a).
This energy shift, present for the duration of the laser pulse, adds a net phase to the polariton given by
\begin{equation}
  \Phi(n,\Delta_j) = \frac{\tau\varepsilon_g}{\hbar}
\end{equation}
which in the far-detuned limit $\abs{\Delta_j}\gg\Omega_0$ reduces to
\begin{equation}\label{eq:TotalPhase}
  \Phi(n,\Delta_j)\simeq \frac{\tau\Omega_0^2}{4\Delta_j} = D\frac{n}{\Delta_j},
\end{equation}
where
\begin{equation}\label{eq:D}
  D = \frac{|\mu|^2F_\mathrm{DW}\Eph\cos^2(\theta)}{2\hbar^2c\nD\varepsilon_0\Aeff}.
\end{equation}
In typical experiments using a high-NA free-space objective to focus on a single NV center through a planar, (100)-oriented, diamond surface, $D/2\pi\approx\!10$~\kilo\hertz\ , so the accumulated phase per photon is only $D/\Delta_j\approx\!10^{-5}$~\radian\ for typical detunings in the \giga\hertz\ range.
Nonetheless, an optical pulse with power $\approx\!1$~\micro\watt\ and duration $\approx\!1$~\micro\second\ contains $\approx\!10^6$ photons, so we can still obtain an observable signal from the total accumulated phase.

\subsection{The Faraday and optical Stark effects}

In order to obtain expressions for the Faraday and optical Stark effects using this model, we need to resolve the resulting polariton state into its spin and optical components.
For that purpose, we calculate the reduced density matrices
\begin{equation}
  \begin{sistema}
    \hat{\rho}_\mathrm{light} = \Tr_\mathrm{spin}(\hat{\rho}) \\
    \hat{\rho}_\mathrm{spin} = \Tr_\mathrm{light}(\hat{\rho})
  \end{sistema}
\end{equation}
in terms of the full density matrix $\hat{\rho}$ for polariton states, which we derive below.
Whereas the polariton states are naturally written in terms of the Fock basis of photon number states, the laser field is best described by an optical coherent state, $\ket{\alpha}$, defined by
\begin{equation}
  \hat{a}\ket{\alpha} = \alpha\ket{\alpha}.
\end{equation}
The coherent state can be expanded in the Fock basis using the relation
\begin{equation}
  \ket{\alpha} = e^{-\frac{\abs{\alpha}^2}{2}}\sum_n\frac{\alpha^n}{\sqrt{n!}}\ket{n},
\end{equation}
which describes a Poisson distribution of Fock states, characterized by mean photon number $\langle n\rangle=\abs{\alpha}^2$ and with uncertainty $\Delta n=\abs{\alpha}=\sqrt{\langle n\rangle}$.
An initial polariton state described by
\begin{equation}
  \ket{\Psi_0} = \left(\sum_j\beta_j\ket{g_j}\right)\otimes\ket{\alpha}
\end{equation}
therefore evolves to the state
\begin{equation}
  \ket{\Psi} = \sum_j\beta_j\:e^{-\frac{\abs{\alpha}^2}{2}}\sum_n\frac{\alpha^n}{\sqrt{n!}}\:e^{i\Phi(n,\Delta_j)}\ket{g_j}\otimes\ket{n}
\end{equation}
after an interaction involving $n$ photons.
Using \cref{eq:TotalPhase} in the limit $\abs{\Delta_j}\gg\Omega_0$ we recast this as
\begin{align}
  \ket{\Psi} & = \sum_j\beta_j\:e^{-\frac{\abs{\alpha}^2}{2}}\sum_n\frac{\left(\alpha e^{i\phi_j}\right)^n}{\sqrt{n!}}\ket{g_j}\otimes\ket{n} \\
   & = \sum_j\beta_j\ket{g_j}\otimes\ket{\alpha e^{i\phi_j}},
\end{align}
where $\phi_j = D/\Delta_j$ is the phase per photon accumulated by the state $\ket{g_j}\otimes\ket{\alpha}$.
The full density matrix of the resulting spin-light system is then given by $\rho = \ket{\Psi}\bra{\Psi}$.

\begin{figure}
\begin{center}
\includegraphics[width=\textwidth]{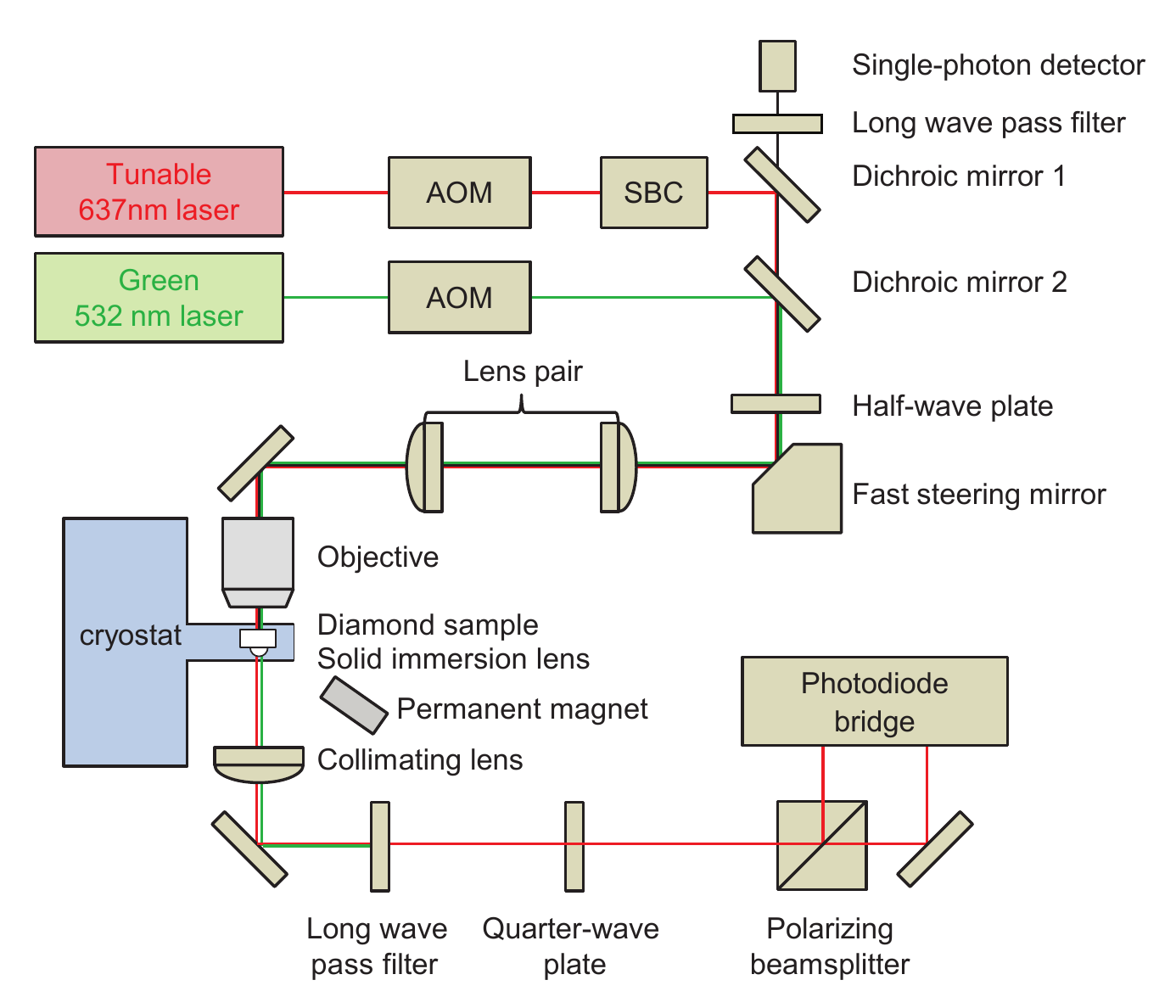}
  \caption[Measurement setup]{\label{fig:BuckleyExpt}
  \textbf{Measurement setup}. Schematic of the experimental setup used to measure Faraday and optical Stark effects.  A tunable laser near the NV-center ZPL at \unit{637}{\nano\metre} provides the coherent optical pulses.  A second laser at \unit{532}{\nano\metre} is used to initialize the NV-center spin and charge state. [AOM: Acousto-optic modulator; SBC: Soleil-Babinet compensator]. Adapted from Ref.~\cite{Buckley2010} and reprinted with permission from AAAS.}
\end{center}
\end{figure}

We first consider the Faraday effect, which describes the observable properties of the laser light following the interaction.
The reduced density matrix for the optical field is readily evaluated as
\begin{align}
  \hat{\rho}_\mathrm{light} & = \sum_k\bra{g_k}\hat{\rho}\ket{g_k} \\
  & = \sum_j\Abs{\beta_j}^2\ket{\alpha e^{i\phi_j}}\bra{\alpha e^{i\phi_j}}.
\end{align}
Thus the optical field is in the state $\ket{\alpha e^{i\phi_j}}$ with a probability $\Abs{\beta_j}^2$ equal to the initial occupation probability of the spin state $\ket{g_j}$.
The observable quantity in this case is the sinusoidal phase of the electric field, which for a coherent state $\alpha = \Abs{\alpha}e^{i\gamma}$ has an expectation value given by
\begin{equation}
  \langle \hat{E}(\vec{x},t)\rangle_\alpha = -\sqrt{2}\mathcal{E}_0\abs{\alpha}\vec{u}(\vec{x})\sin(\omega t-\gamma),
\end{equation}
where $\vec{u}(\vec{x})$ describes the spatial mode and $\mathcal{E}_0$ is the vacuum electric field \cite{Gerry2005}.
The complex phase of the coherent state $\ket{\alpha}$ is therefore reflected as the phase of the electric field.
In the experiment by Buckley \textit{et al.} \cite{Buckley2010}, only one linear polarization of light is coupled to the transition $j$.
Its phase is shifted relative to the non-interacting polarization state by an amount $\phi_j$, which rotates the linear polarization angle of the transmitted light.
\Cref{fig:BuckleyExpt} shows a schematic of the experimental setup.

The experiment is performed in the intermediate-strain regime ($2\delta\approx\unit{17}{\giga\hertz}$) where the excited-state orbitals are energetically separated and can be individually addressed.
The approximate level structure of the ground state and lower-branch excited state is shown in \cref{fig:LightMatterCoupling}(b); a relatively large axial magnetic field of $B_z=1920$~G ensures that the $\hat{S}_z$ eigenstates are a good spin basis for the excited state, however the spin-spin and spin-orbit interactions shift the energies relative to the ground state as shown by \cref{eq:HL}.
Thus, the optical resonance for different spin sublevels occur at different frequencies, with the $m_s=-1$ transition roughly \unit{3}{\giga\hertz} lower in frequency than the $m_s=0$ transition.

We define the Faraday phase $\PhiF$ as the difference in phase between the $m_s\!=\!0$ and $m_s\!=\!-1$ spin states, given by
\begin{equation}\label{eq:FE}
  \PhiF = \phi_0-\phi_{-1} = D\left(\frac{1}{\Delta_0} - \frac{1}{\Delta_{-1}}\right) = -D\frac{\omega_s}{\Delta_0\Delta_{-1}},
\end{equation}
where $\omega_s = (E_{-1}-E_{0})/\hbar$ is the frequency spacing between the
resonances.

Similarly, the reduced density matrix for the spin is given by
\begin{align}\label{eq:SpinPart}
  \hat{\rho}_\mathrm{spin} & = \bra{\alpha}\hat{\rho}\ket{\alpha} \\
  & = \sum_{j,k}\beta_k^\ast\beta_j\exp\left[-\abs{\alpha}^2\left(2-e^{i\phi_j} - e^{-i\phi_k}\right)\right]\ket{g_j}\bra{g_k},
\end{align}
where we have used the identity
\begin{equation}
  \braket{\alpha}{\alpha^\prime} = \exp\left[-\frac{1}{2}\left(\abs{\alpha}^2+\abs{\alpha^\prime}^2-2\alpha^\ast\alpha^\prime\right)\right].
\end{equation}
Since $\phi_j\ll 1$, we can approximate
\begin{equation}
  \hat{\rho}_\mathrm{spin}\simeq \sum_{j,k}\beta_k^\ast\beta_j\:e^{i\langle n\rangle(\phi_j-\phi_k)}\ket{g_j}\bra{g_k},
\end{equation}
from which we identify the effective spin states
\begin{equation}
  \ket{\mathrm{spin}} = \sum_j\beta_j\:e^{i\langle n\rangle\phi_j}\ket{g_j} = \sum_j\beta_j\:e^{i\frac{\tau\Omega_0^2}{4\Delta_j}},
\end{equation}
such that $\hat{\rho}_\mathrm{spin} = \ket{\mathrm{spin}}\bra{\mathrm{spin}}$.
Physically, this shows that the spin states acquire relative phases due to their different detunings from the light field, producing an effective spin rotation.
In the experiment \cite{Buckley2010}, this relative optical-Stark-effect phase is directly proportional to the corresponding Faraday-effect phase through the photon number:
\begin{equation}
  \Phi_\mathrm{OSE} = n\PhiF.
\end{equation}
The optical field in the experiment consists of two polarization modes, each with photon number $n$, of which only one is coupled to the NV-center optical transitions, so the total laser power is given by $P_\mathrm{L} = 2n\Eph/\tau$, and the corresponding optical Stark frequency shift is
\begin{equation}\label{eq:Sose}
   \Sose = \frac{\Phi_\mathrm{OSE}}{2\pi\tau} =\frac{P_\mathrm{L}}{4\pi\Eph}\PhiF.
\end{equation}
This proportionality in the far-detuned regime allows the two measurements to be shown together on the same graph as in \cref{fig:LightMatterCoupling}(c).

Although the expressions above were derived assuming the limit of large detuning ($\abs{\Delta_j}\gg\Omega_0$), the full expressions across the absorption resonance are known from other arguments.
The Faraday effect results from the real part of the frequency-dependent refractive index of the atomic transition near an absorption resonance.
As a consequence of the Kramers-Kronig relation between the refractive index's real and imaginary parts, the full Faraday effect lineshape is known to be an odd Lorentzian of the form
\begin{equation}\label{eq:OddLorentzian}
  \phi_j = \frac{\mathcal{F}_j\Delta_j}{\Delta_j^2+\Gamma_j^2},
\end{equation}
where $\Gamma_j$ is the width of absorption resonance $j$ and $\mathcal{F}_j$ is the Faraday amplitude.
By comparing this expression with \eqnref{eq:FE} in the far-detuned limit we see that the constant $D$ takes the place of the Faraday amplitude $\mathcal{F}$.
Likewise, the shift in the Larmor precession rate due to the optical Stark effect is a direct consequence of the polariton energy shift of \eqnref{eq:EnergyShift}, and so is given across
all detunings by
\begin{equation}\label{eq:StarkShift}
  S_j = \frac{\Delta_j}{4\pi}\left[\sqrt{1+\frac{\Omega_0^2}{\Delta_j^2}}-1\right].
\end{equation}
In comparing measurements to these expressions, we can extract experimental values for $\mathcal{F}_j$, $\Gamma_j$, and $\Omega_0$ for the appropriate transitions.
For the data in \cref{fig:LightMatterCoupling}(c) from Ref. \cite{Buckley2010}, we obtain $\mathcal{F}_0=2\pi\times 6.9$~\micro\radian\usk\giga\hertz, $\Gamma_0=2\pi\times140$~\mega\hertz, $\mathcal{F}_{-1}=2\pi\times 7.6$~\micro\radian\usk\giga\hertz, $\Gamma_{-1}=2\pi\times300$~\mega\hertz\, and $\Omega_0\!=\!2\pi\times70$~\mega\hertz.
The asymmetry in the curve mainly results from the different absorption widths for the $m_s=0$ and $m_s=-1$ transitions.

\subsection{Discussion and implications}

The preceding derivation illustrates how coherent light-matter interactions give rise to observable spin-dependent optical phase shifts (the Faraday effect) and coherent, optical-power-dependent spin rotations (the optical Stark effect).
In principle, the Faraday effect provides a means to measure the spin state nondestructively, \textit{i.e.}, without exciting the optical transition and re-initializing the state.
This is possible since the absorption resonance has a Lorentzian lineshape, varying as $1/\Delta^2$ for large $\Delta$, whereas the Faraday phase shift is an odd Lorentzian, varying as $1/\abs{\Delta}$.
Nondestructive measurements are important for certain applications in quantum information processing, and similar dispersive measurements are used extensively in the circuit quantum electrodynamics paradigm of superconducting qubits \cite{Blais2004}.
In practice, the Faraday phase shifts on the order of $10^{-5}$~rad are too small to allow high-fidelity, non-destructive measurements of individual NV centers without an optical cavity to amplify the interaction.
Although it remains a challenge to fabricate nanophotonic optical cavities containing NV centers while maintaining stable optical transitions, such a platform has recently been achieved for silicon-vacancy (SiV) centers in diamond \cite{Sipahigil2016}, where dispersive interactions analogous to those we have discussed above can also serve to mediate interactions between two SiV spins within the same cavity \cite{Evans2018}.

The optical Stark effect, meanwhile, provides a means to perform operations on a spin qubit using light rather than microwaves, which can allow addressing of individual qubits within optical networks.
With enhanced interactions from an optical cavity, the optical Stark effect can provide a means for generating spin-photon entanglement or quantum operations between remote spins.
Whereas the spin rotations that result from the optical Stark shifts in a level structure like \cref{fig:LightMatterCoupling}(b) generate precession about the qubit's polar axis, variations in the energy level structure and experimental design can enable rotations about arbitrary axes on the Bloch sphere, in addition to general protocols for qubit readout and initialization \cite{Yale2013}, as we discuss in the next section.


\section{All-optical coherent spin control\label{sec:CPT-SRT}}

In the previous section, the optical Stark effect\,---\,viewed as the spin-like component of the coherent polariton dynamics as in \cref{eq:SpinPart}\,---\,manifests as a relative energy shift between spin sublevels, with no change in the spin eigenstates.
This is analogous to the application of a magnetic field along the defect's symmetry axis.
When treating two of the triplet spin sublevels as a qubit, this amounts to a light-induced rotation about the $z$ axis in the Bloch sphere.
In order to achieve arbitrary unitary operations on a qubit, however, rotations about two noncollinear axes are required.
One can therefore ask if it is possible to realize optical Stark effects that perturb the ground-state Hamiltonian in more complex ways, \textit{e.g.}, to generate an effective magnetic field pointing along $x$ or $y$.
Indeed, this is possible if one can engineer the electronic structure and optical transition diagram to enable light-induced mixing of the spin eigenstates.

\begin{figure}
\begin{center}
\includegraphics[]{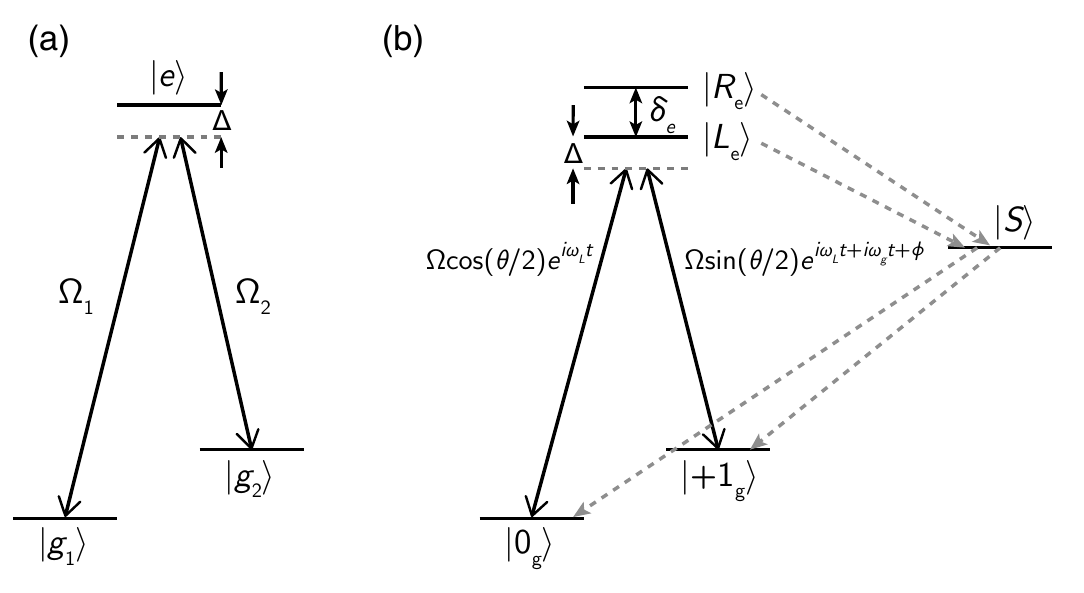}
  \caption[Lambda system]{\label{fig:CPT}
  \textbf{Physics of $\Lambda$ configurations}. (a) Three levels arranged in a $\Lambda$ configuration. (b) Realization of a $\Lambda$ system for the NV center from an excited-state avoided level crossing.}
\end{center}
\end{figure}

Such mixing occurs naturally in a level configuration known as a lambda ($\Lambda$) system, where two lower-energy states (the qubit manifold) couple coherently to a single excited state, as shown in \cref{fig:CPT}(a).
Lambda configurations occur in a variety of quantum systems including atoms \cite{Alzetta1976,Gray1978}, trapped ions \cite{Wineland1999}, quantum dots \cite{Xu2008}, and superconducting qubits \cite{Kelly2010}.
As we will show below, the concept of optical Stark rotations as applied to a $\Lambda$ system can be extended to realize arbitrary qubit operations; in this context they are known as stimulated Raman transitions.
Furthermore, the $\Lambda$ configuration is the basis for many well-known effects in quantum optics, including coherent population trapping (CPT) \cite{Alzetta1976}, electromagnetic induced transparency \cite{Boller1991}, slow light \cite{Budker1999}, atomic clocks \cite{Vanier2005}, and spin-photon entanglement \cite{Togan2010}.

\subsection{Dark states and coherent population trapping}

The essential feature of a $\Lambda$ system is the appearance of ``dark resonances'' that occur when two light fields coherently drive both transitions to the excited state.
When the light fields are tuned such that their frequency difference exactly matches the resonance frequency of the ground-state sublevels, the atom is no longer pumped to the excited state and therefore becomes dark.
This phenomenon can be simply understood from the following argument \cite{Cohen-Tannoudji2011}.
If the atom is initially in a superposition of ground states,
\begin{equation}\label{eq:GSsuper}
  \ket{\psi(t=0)}=c_1\ket{g_1}+c_2\ket{g_2},
\end{equation}
and it interacts with two laser fields characterized by instantaneous Rabi frequencies (here assumed to be complex quantities),
\begin{equation}
  \Omega_i=\frac{\vec{\mu}_i\cdot\vec{E}_i}{\hbar},
\end{equation}
then the amplitude for a transition to occur from state $\ket{g_i}$ to the excited state, $\ket{e}$, is proportional to the product $c_i\Omega_i$.
If a ground-state superposition $\ket{\psi}$ exists such that
\begin{equation} \label{eq:DarkState}
  c_1\Omega_1+c_2\Omega_2=0,
\end{equation}
then the amplitudes for the transitions from both ground states interfere destructively, and the atom cannot be excited.
This is called a \emph{dark state}.
Since both the probability amplitudes $c_i$ and the electric field amplitudes $\vec{E}_i$ are functions of time, the atom is not guaranteed to stay in a dark state indefinitely; however, it is straightforward to show that the condition of \cref{eq:DarkState} is maintained continuously if
\begin{equation}\label{eq:CPTcriterion}
  \varepsilon_2-\varepsilon_1 = \hbar(\omega_1-\omega_2),
\end{equation}
where $\varepsilon_i$ and $\omega_i$ are the ground-state energies and laser frequencies, respectively, \textit{i.e.}, if the detuning of the two light fields matches the ground-state energy splitting.

The existence of a persistent dark state results in the phenomena of CPT and electromagnetic induced transparency.
Starting from an arbitrary ground-state configuration and subject to light fields satisfying \cref{eq:CPTcriterion}, the atom is transiently excited and relaxes until it is trapped in the dark state and no longer interacts with the optical fields.
One can think of this dissipative process as a generalization of traditional optical pumping, \textit{i.e.}, where only one arm of the $\Lambda$ system is driven.
Intuitively, if only transition 1 is driven, the system will quickly relax into a steady state with $\ket{g_2}$ fully populated, uncoupled to the optical field.
This scenario is a special case of \cref{eq:DarkState} with $\Omega_2=0$, where the dark state is $\ket{D}=\ket{g_2}$.
In fact, a dark state satisfying \cref{eq:DarkState} is guaranteed to exist for any values of $\Omega_i$, and there will always be a corresponding \emph{bright state}, $\ket{B}$, that is orthogonal to $\ket{D}$ and couples maximally to the optical field.
Thus, by choosing the amplitude and phases of the optical fields that define $\Omega_i$, one can initialize the system into an arbitrary superposition of qubit ground states.

While the CPT process is necessarily dissipative (\textit{i.e.}, non-unitary), coherent evolution in the ground state can be achieved using dispersive interactions in analogy with the optical Stark effect.
When the optical fields satisfying \cref{eq:CPTcriterion} are simultaneously detuned from the resonance condition with $\ket{e}$ as shown in \cref{fig:CPT}(a), the resulting light shift occurs only for the state $\ket{B}$ and not $\ket{D}$.
In the qubit manifold, this manifests as a light-induced rotation about the axis pointing from $\ket{D}$ to $\ket{B}$ in the Bloch sphere.
The axis can be chosen arbitrarily, including configurations on the equator when $\abs{\Omega_1}=\abs{\Omega_2}$ that result in complete population transfer between the qubit eigenstates.
The effect in this context is usually known as stimulated Raman transitions (SRTs), drawing inspiration from an alternative picture of the process in terms of virtual transitions through the excited state.
However, it is important to understand that SRTs are a dispersive effect that do not involve absorption.
Again, whereas CPT varies with $1/\Delta^2$, where $\Delta$ is the detuning from the optical resonance(s), the effective Rabi frequency of SRTs scales with $1/\Delta$, so it can be substantial even when absorption is negligible.

On a practical note, it is important to recognize that the condition to have a dark state can only be sustained if the two optical fields have a deterministic phase relationship.
If the fields are derived from different lasers, they must be frequency and phase stabilized to a suitable reference.
Alternatively, if the required frequency difference occurs in the radiofrequency or microwave spectrum, the two fields can be derived from a single laser using an optical modulator to generate frequency sidebands.
This is often the easiest approach, and it is the one adopted by Ref.~\cite{Yale2013}.

\subsection{Forming a $\Lambda$ system from the NV center}

NV centers in diamond have long been known to exhibit electromagnetic-induced transparency and CPT \cite{Harley1984,Reddy1987,Hemmer2001,Santori2006}, evidence that $\Lambda$ configurations can be realized under certain conditions.
At zero strain and zero magnetic field, the spin-orbit eigenstates $\ket{A_1}$ and $\ket{A_2}$ are equal superpositions of the $m_s=\pm1$ spin eigenstates, $\ket{\pm1_g}$, with circular-polarization optical selection rules that facilitate the generation of spin-photon entanglement \cite{Togan2010} and CPT in the $\{\ket{+1_g},\ket{-1_g}\}$ ground-state subspace \cite{Togan2011}.
However, it is often more convenient to work with a ground-state qubit defined in a manifold including the $m_s=0$ sublevel, $\ket{0_g}$, since this state is naturally prepared by off-resonant optical pumping, and at low temperature it features optical cycling transitions that facilitate robust, high-fidelity readout \cite{Robledo2011a}.

As is apparent from the spin-spin terms in the excited-state Hamiltonian, Eq. (4), and the approximate spin-triplet representations in the high-strain regime, \cref{eq:HL,eq:HU}, the excited-state $m_s=0$ states are weakly admixed with $m_s=\pm1$ by the spin-spin parameter $\Delta_2$.
However, this parameter is rather small ($\Delta_2=\unit{150}{\mega\hertz}$ \cite{Bassett2014}), so the mixing only becomes apparent near an avoided level crossing, when the $m_s=0$ sublevel becomes nearly degenerate with $m_s=+1$ or $m_s=-1$.
Such a situation is depicted in \cref{fig:CPT}(b), where the applied magnetic field is tuned such that a crossing occurs between the $m_s=+1$ and $m_s=0$ spin sublevels of the lower orbital branch, $\{\ket{+1_e},\ket{0_e}\}$.
(This particular crossing can only occur when the strain is relatively small, since for large transverse perturbations the $\ket{+1_e}$ state is higher in energy than $\ket{0_e}$ even when $B=0$; see \cref{fig:NVstructure}(b).)
At the closest approach, the anticrossing levels are separated by an energy $\delta_e\approx\Delta_2$, and the eigenstates become
\begin{subequations}\label{eq:RLbasis}
  \begin{align}
  \ket{R_e} & = \frac{1}{\sqrt{2}}(\ket{0_e}+\ket{+1_e)} \\
  \ket{L_e} & = -\frac{1}{\sqrt{2}}(\ket{0_e}-\ket{+1_e)}.
  \end{align}
\end{subequations}
Either of these states can serve as the upper state of a $\Lambda$ system connecting the $\{\ket{+1_e},\ket{0_e}\}$ qubit states.

Yale \textit{et al.} \cite{Yale2013} explored this situation by tuning to an excited-state avoided level crossing as shown in \cref{fig:CPT}(b) and modulating a tunable laser near \unit{637}{\nano\metre} using an electro-optic phase modulator in order to generate sidebands separated by the ground-state resonance frequency, $\omega_\mathrm{gs}$.
This also allows for control of the relative phase between the two optical fields and their relative amplitude through the power and phase of the microwave signal applied to the modulator.
These parameters determine the azimuthal ($\phi$) and polar ($\theta$) angles of the dark state formed in the ground-state Bloch sphere.

\subsection{All-optical initialization, control, and readout}

To describe the dynamics of the NV-center spin under optical excitation as shown in \cref{fig:CPT}(b), we construct a model including five energy levels: two out of the three ground-state levels $|0_g\rangle$, $|+1_g\rangle$, the two mixed excited states $|L_{e}\rangle$ and $|R_{e}\rangle$, as well as the intermediate singlet $|S\rangle$, which plays a role in mediating unintentional ISC transitions that cause dissipation.
The Hamiltonian, in the rotating frame, for the subspace spanned by these five basis states can be expressed as
\begin{equation}\label{eq:Hcpt}
H =
 \begin{pmatrix}
  \Delta_L & 0 & \Omega \cos(\theta/2) & \Omega \cos(\theta/2) & 0 \\
  0 & \Delta_L & \Omega \sin(\theta/2)e^{i\phi}  & -\Omega \sin(\theta/2)e^{i\phi}   & 0 \\
  \Omega \cos(\theta/2) & \Omega \sin(\theta/2)e^{-i\phi}  & 0  & 0  & 0 \\
  \Omega \cos(\theta/2) & -\Omega \sin(\theta/2)e^{-i\phi} & 0  & -\delta_{e}  & 0 \\
  0 & 0 & 0 & 0 & \epsilon_S
 \end{pmatrix}
\end{equation}
where the ordering of the states is $\{|+1_g\rangle,|0_g\rangle,|R_{e}\rangle,|L_{e}\rangle, |S\rangle\}$, $\Delta_L$ is the detuning of the laser frequency ($\omega_L$)  from resonance to the $|R_{e}\rangle$ $\Lambda$ system, $\delta_{e}$ is the separation of the excited state levels, $\Omega$ is the optical Rabi frequency, $\phi$ is the relative phase between the two coherent light fields, and $\tan(\theta/2)$ is the relative amplitude between the driving fields.

The time evolution of the system includes both coherent and dissipative processes.
These can be captured using the Lindblad master equation,
\begin{equation}\label{eq:LindbladME}
\dot\rho = i\left[\rho,H\right]
+ \sum_{\alpha, \alpha'} \Gamma_{\alpha \alpha'}  \left(\sigma_{\alpha'\alpha}\rho \sigma_{\alpha\alpha'}
                 -\frac{1}{2}\sigma_{\alpha\alpha}\rho  -\frac{1}{2}\rho\sigma_{\alpha\alpha} \right)
\equiv W\rho.
\end{equation}
The first term describes unitary evolution of the density matrix due to the Hamiltonian of \cref{eq:Hcpt}, whereas the second term captures dissipative processes, with the Lindblad operators ${\sigma_{\alpha\alpha} = |\alpha\rangle\langle\alpha|} = \sigma_{\alpha'\alpha}^\dagger \sigma_{\alpha'\alpha}$ and ${\sigma_{\alpha'\alpha} = \sigma_{\alpha\alpha'}^\dagger = |\alpha'\rangle\langle\alpha|}$.
%
For $n=5$ levels, the density matrix $\rho$ is a Hermitian 5x5 matrix and can be described by $n^2=25$ real parameters ($n^2-1=24$ including the normalization condition $\mathrm{Tr}\left(\rho\right) =1$).
The superoperator $W$ can thus be viewed as a 25x25 matrix with rank 24.

The Lindblad operators describe incoherent, spontaneous transitions between states.
We denote the decay rate from the excited states (${E=L_e,R_e}$) to the ground states ($G=0,1$) with ${\Gamma = \Gamma_{E,G_g}}$, the rate for ISC from the excited states to the singlet ${\Gamma_i = \Gamma_{E,S}}$, and the inverse ISC rate from $|S\rangle$ to one of the ground state levels as ${\Gamma_i' = \Gamma_{S,G_{g}}}$.
The spin relaxation rate in the ground state is $\Gamma_1 =1/T_1$. 
At $T\approx 10\,{\rm K}$, the thermal frequency $k_BT/h\approx200\,{\rm GHz}$ exceeds the relevant NV level splittings $\approx 1\,{\rm GHz}$ by orders of magnitude, and therefore $\Gamma_{+1_g,0_g}=\Gamma_{0_g,+1_g}= \Gamma_1/2$.
Pure dephasing between the two ground state levels is approximated by adding a term ${\gamma = 1/T_2 =\Gamma_{0_g,0_g}}$.
All other rates are set to zero.

The state of the system after optical excitation during time $t$ is obtained by integrating \cref{eq:LindbladME},
\begin{equation}
\rho(t) = e^{W t}\rho(0),\label{eq:dynamics}
\end{equation}
where the initial state, $\rho(0)$, is typically one of the ground states.
\Cref{eq:dynamics} admits simple analytical solutions only for special cases, so in general we simulate the dynamics numerically.
Depending on the parameters, this model can describe both CPT and SRT.
In the idealized case ${\Gamma_1 = \gamma = \Gamma_i =0}$, and with only one of the excited
levels included, the system reduces to the three-level $\Lambda$ system of \cref{fig:CPT}(a), and the stationary state $\bar \rho$ in the long-time limit $t\gg 1/\Gamma$ obtained from ${\dot\rho=0}$ as the null space of $W$ is the dark state:
\begin{equation}\label{eq:NVdarkstates}
|D\rangle = \cos(\theta/2) |0_g\rangle -\exp(\mp i\phi)\sin(\theta/2) |+1_g\rangle
\end{equation}
where the upper (lower) sign holds for the single excited state level being ${E=R}$ (${E=L}$).

With realistic parameters, the evolution is not so simple, since the ISC and spin decoherence tend to disspate the system away from the ideal dark state.
Furthermore, we notice from \cref{eq:NVdarkstates} that the dark states corresponding to the different excited states $\ket{L_e}$ and $\ket{R_e}$ have opposite phases.
When these states lie on the equator ($\theta=0$), they are orthogonal, such that the dark states from one $\Lambda$ system is actually the bright state from the other.
Since the separation between these states is small ($\delta_e/h\sim\unit{180}{\mega\hertz}$ in Ref.~\cite{Yale2013}), there exists a tradeoff between the speed of the operations, set by the laser power, and the competition between these two orthogonal $\Lambda$ systems, which becomes more prevalent as the laser power increases.

In any case, the time-dynamics of the Bloch vector representing the qubit density matrix can be obtained from
\begin{equation}
\mathbf{b}(t) = \mathrm{Tr} \left(\boldsymbol{\sigma}\rho(t)\right),
\end{equation}
where the components of $\boldsymbol{\sigma}$ are the Pauli matrices
in the ground-state subspace,
\begin{eqnarray}
\sigma_\mathrm{x} &=& |+1_g\rangle\langle 0_g| + |0_g\rangle\langle +1_g|, \\
\sigma_\mathrm{y} &=&  i (|+1_g\rangle\langle 0_g| - |0_g\rangle\langle +1_g|), \\
\sigma_\mathrm{z} &=& |0_g\rangle\langle 0_g| - |+1_g\rangle\langle +1_g|.
\end{eqnarray}
The fidelity of an operation can be calculated by comparing the final density matrix to a target state, \textit{e.g.} for initialization \textit{via} CPT in the dark state $\ket{D}$,
\begin{equation}
F(t) = \langle D| \rho(t) |D\rangle.
\end{equation}

\begin{figure}
\begin{center}
\includegraphics[width=\textwidth]{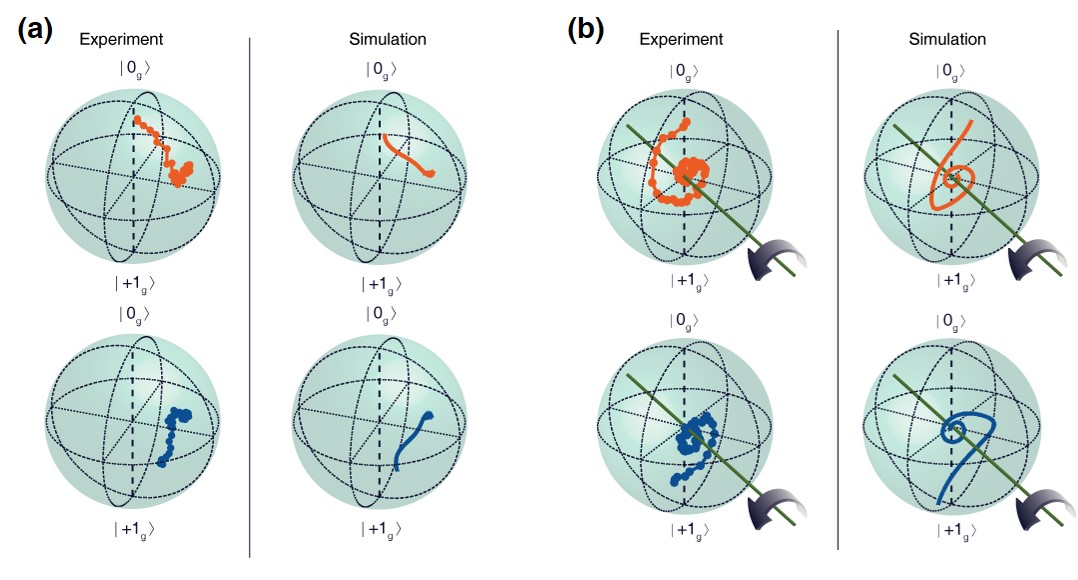}
  \caption[Lambda system]{\label{fig:CPT-SRT}
  \textbf{All-optical control \textit{via} coherent dark states}. Experiments (points) and simulations (curves) of quantum dynamics in the NV-center ground state driven by optical pulses designed to achieve CPT (a) and SRT (b). Orange (top) and blue (bottom) trajectories correspond to situations where the initial state is $\ket{0_g}$ or $\ket{+1_g}$, respectively. Adapted from Ref.~\cite{Yale2013} and reprinted with permission from the National Academy of Sciences.}
\end{center}
\end{figure}

\Cref{fig:CPT-SRT} shows examples of experimental CPT and SRT trajectories from Yale \textit{et al.} \cite{Yale2013} alongside simulations performed using this model.
The measurements (points) are acquired by performing Bayesian quantum state tomography to reconstruct the state vector from experiments where the NV-center spin is repeatedly initialized, subjected to a particular optical pulse, and then measured in one of three orthogonal bases.
In addition to arbitrary-basis initialization and coherent control \textit{via} CPT and SRT, respectively, Yale \textit{et al.} \cite{Yale2013} also demonstrated how the intrinsic fluorescence contrast between the bright and dark state can be used to perform projective readout of the spin state in an arbitrary basis.
It is thus possible to perform full quantum operations, for example Rabi, Ramsey, or Hahn-echo spin coherence measurements, using light fields alone.
Crucially, these methods do not rely on the NV center's intrinsic level structure and spin-dependent ISC dynamics; they can be adapted to any system where a $\Lambda$ configuration can be realized through tuning of external electric or magnetic fields.
Indeed, the methods have recently been adapted to study the quantum properties of spin defects that do not exhibit an ISC, for example the negatively-charged SiV in diamond \cite{Rogers2014a,Pingault2014} and transition-metal impurities in silicon carbide \cite{Koehl2017}.


\section{Ultrafast control\label{sec:Ultrafast}}

The versatile concepts of light-matter coupling presented in \cref{sec:LightMatterCoupling,sec:CPT-SRT} underlie many applications in quantum optics and quantum information science.
In particular, dispersive effects such as the Faraday phase shift, the optical Stark shift, and stimulated Raman transitions provide a means to perform coherent quantum operations on individual spins and to generate quantum correlations between light and matter.
However, practical limitations mean they are not always the most efficient method to control solid-state defects.
Although the technique is all-optical in the sense that only light fields interact with the spin, generation of the requisite phase-locked optical fields demands stable, tunable laser sources, optical modulators, and corresponding microwave equipment.
Moreover, the CPT and SRT trajectories shown in \cref{fig:CPT-SRT} exhibit several drawbacks of this technique as applied to the NV center specifically.
The CPT trajectories do not terminate on the surface of the Bloch sphere, indicating a partially mixed initialized state, and the SRT trajectories rapidly spiral inwards towards a totally mixed state at the Bloch-sphere center.
These nonidealities result from various experimental factors such as laser noise and spectral drift of the NV-center optical resonances, and from intrinsic properties of the NV center.
One key limitation is the small spin-spin coupling parameter, $\Delta_2/h \approx\unit{150}{\mega\hertz}$, responsible for the excited-state anticrossing that forms a pair of $\Lambda$ systems for the $\{\ket{0},\ket{+1}\}$ spin sublevels as in \cref{fig:CPT}(b).
Since the bright state from one $\Lambda$ system is the dark state for the other, competing dynamics between the two $\Lambda$ systems limit the fidelity of CPT initialization and add decoherence to SRT operations.
This dual-$\Lambda$ configuration also limits the effective speed of SRT operations (\textit{i.e.}. the ground-state Rabi frequency, $\Omega_g$) such that $\hbar\Omega_g\ll\Delta_2$.
For the NV center, the practical limit is $\Omega_g/2\pi \approx \unit{10}{\mega\hertz}$, whereas traditional microwave control of the ground state can facilitate high-fidelity operations at Rabi frequencies approaching \unit{1}{\giga\hertz} \cite{Fuchs2009}.

\subsection{Quantum control with ultrafast optical pulses}

\begin{figure}
\begin{center}
\includegraphics[width=\textwidth]{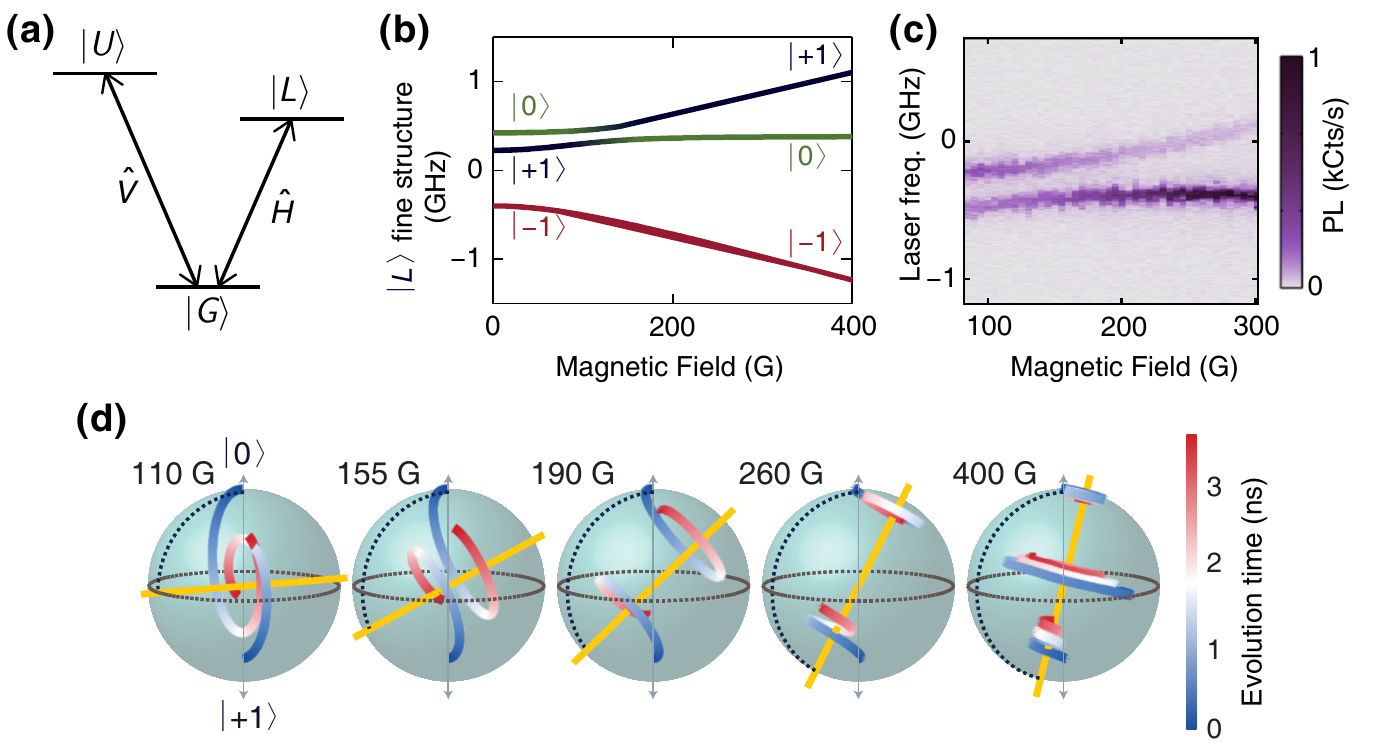}
  \caption[Ultrafast control]{\label{fig:UFcontrol}
  \textbf{Coherent spin control with ultrafast pulses}.  (a) Orbital structure of the NV center at intermediate-to-high transverse strain. (b) Fine structure as a function of axial magnetic field in the $\ket{L}$ orbital branch when $\delta=\unit{6.7}{\giga\hertz}$. (c) PL excitation spectroscopy as a function of axial magnetic field, showing the level anticrossing between $\ket{L,0}$ and $\ket{L,+1}$ around $B=110$~G. (d) Trajectories of the ground-state spin qubit as a function evolution time between two optical pulses, for different settings of the magnetic field. Adapted from Ref.~\cite{Bassett2014} and reprinted with permission from the AAAS.}
\end{center}
\end{figure}

In this section, we introduce an alternate approach to achieving all-optical quantum control using ultrafast optical pulses that mitigates some of these limitations \cite{Bassett2014}.
This approach abandons the dispersive approximation of negligible optical excitation; rather, we directly leverage dynamics generated by the excited-state Hamiltonian to achieve desired unitary operations on the spin.
\Cref{fig:UFcontrol}(a) shows the NV center's orbital structure in the intermediate-to-high strain regime.
As described in \cref{sec:NVstrain}, transverse strain splits the excited state into two orbital manifolds, each of which are connected to the ground state \textit{via} orthogonal, linear-polarization selection rules.
Whereas previously we considered optical pulses derived from a continuous-wave laser with durations measured in nanoseconds, which can resolve the NV center's gigahertz-scale fine structure, an optical pulse with duration $\lesssim\unit{1}{\pico\second}$ has a bandwidth $\gtrsim\unit{1}{\tera\hertz}$, much larger than the spin-dependent frequency splittings of the ground and excited states.
Such pulses operate on the orbital degrees of freedom only, effectively altering the orbital population instantaneously from the point of view of the spin dynamics.

The orbital Hamiltonian with $h=1$ in the $\{ |G\rangle$,$|X\rangle$,$|Y\rangle\}$ basis with a strain $\delta$ in direction $\alpha_{S}$ is given by
\begin{equation}
        \mathbf{H_{orb}} =
        \begin{pmatrix}
        0 & 0 & 0 \\
        0 & f_{0} - \frac{\delta}{2}\cos(\alpha_{S}) & \frac{\delta}{2}\sin(\alpha_{S}) \\
        0 & \frac{\delta}{2}\sin(\alpha_{S}) & f_{0} + \frac{\delta}{2}\cos(\alpha_{S})
        \end{pmatrix},
\end{equation}
where $f_0 = c/\lambda$ is the optical transition frequency.
Each pulse corresponds to a unitary operation on the orbital states, with parameters determined by the pulse intensity, shape, and polarization.
We parameterize the electric field of the optical pulses by
\begin{equation}
  \mathcal{E}(\alpha_E,\beta_E) = \left(
  \begin{array}{c}
     \cos(\alpha_E)\cos(\beta_E) - i\sin(\alpha_E)\sin(\beta_E)  \\
     \sin(\alpha_E)\cos(\beta_E) + i\cos(\alpha_E)\sin(\beta_E)
  \end{array}\right),
\end{equation}
where $\alpha_E$ is the angle of the linearly-polarized component (major axis) in the NV center's $(x,y)$ plane, and $\beta_E\in [-\frac{\pi}{4},\frac{\pi}{4}]$ defines the ellipticity, such that $\beta_E=0$ and $\beta_E=\pm\frac{\pi}{4}$ correspond to linearly and circularly polarized light, respectively.
Using the dipole matrix elements $\bra{X}\hat{y}\ket{G} = -\bra{Y}\hat{x}\ket{G}$ (other combinations vanish), we find that a pulse of polarization $\mathcal{E}(\alpha_E,\beta_E)$ couples $\ket{G}$ to the orbital state
\begin{equation}\label{eq:CoupledES}
  \ket{E} =\left(
  \begin{array}{c}
     0 \\
     -\mathcal{E}_y  \\
     \mathcal{E}_x
  \end{array}\right),
\end{equation}
leaving the orthogonal ES basis state,
\begin{equation}
  \ket{E'} =\left(
  \begin{array}{c}
     0 \\
     \mathcal{E}_x^\ast  \\
     \mathcal{E}_y^\ast
  \end{array}\right),
\end{equation}
unaffected.

In the experiments by Bassett \textit{et al.} \cite{Bassett2014}, pairs of pulses were derived from a single seed laser using beamsplitters and a delay line, so they were nominally identical.
In this case, we can treat the pulses as instantaneous unitary operators parameterized by a rotation angle, $\theta$, and with a relative phase, $\phi$:
\begin{subequations}\label{eq:PulseOperators}
  \begin{align}
  \begin{split}
    U_\mathrm{FP1} & = \ketbra{E'}{E'} + \cos\left(\frac{\theta}{2}\right)\bigl(\ketbra{E}{E}+\ketbra{G}{G}\bigr) \\
        & \qquad + \sin\left(\frac{\theta}{2}\right)\bigl(\ketbra{E}{G}-\ketbra{G}{E}\bigr),
  \end{split} \\
  \begin{split}
    U_\mathrm{FP2} & = \ketbra{E'}{E'} + \cos\left(\frac{\theta}{2}\right)\bigl(\ketbra{E}{E}+\ketbra{G}{G}\bigr) \\
        & \qquad + \sin\left(\frac{\theta}{2}\right)\bigl(e^{i\phi}\ketbra{E}{G}-e^{-i\phi}\ketbra{G}{E}\bigr).
  \end{split}
  \end{align}
\end{subequations}
Between the pulses, the system freely evolves according to the system Hamiltonian.
The evolution can include both unitary and dissipative processes, \textit{e.g.,} following a Lindblad master equation similar to \cref{eq:LindbladME}.

Even though the pulses only act on the orbital degrees of freedom directly, spin-orbit interactions in the excited state naturally induce spin dynamics during the free evolution period.
Depending on the pulse parameters, this scheme can be adapted to probe both orbital and spin dynamics on timescales spanning femtoseconds to nanoseconds, and to realize deterministic control over the spin.
For example, a pair of phase-locked optical pulses can be designed to perform a generalized Ramsey sequence on the three-dimensional orbital Hamiltonian, where the first pulse generates a coherent superposition of ground and excited states that proceeds to evolve, and the second pulse projects the resulting state onto the measurement basis of excited states (which emit PL) and the ground state (which is dark).
This scheme can be adapted to probe orbital coherence between the ground state and excited states or (by tuning the polarization to excite a superposition of $\ket{L}$ and $\ket{U}$) coherence within the excited-state manifold.
Alternatively, by setting $\theta=\pi$, the optical pulses can be designed to achieve full population transfer from $\ket{G}$ to a desired excited state orbital, and \textit{vice versa}.
From the point of view of the spin, this manifests as an instantaneous change in the Hamiltonian.
For a pair of such pulses that populates and subsequently depopulates the excited state after a time, $t$, the excited-state evolution generates a deterministic unitary operation on the spin.

\subsection{Applications}

This novel approach to generating coherent spin rotations by utilizing free evolution in the excited state has several applications.
As a time-domain spectroscopy technique, measurements of the spin dynamics that result from pairs of optical pulses provide the means to map an arbitrary excited-state Hamiltonian.
The technique is termed \emph{time-domain quantum tomography} (TDQT).
In contrast to frequency-domain spectroscopies which typically yield only the Hamiltonian eigenvalues, TDQT yields both the eigenvalues and eigenvectors, from which it is possible to construct the full Hamiltonian matrix.
TDQT also provides time-domain information about various non-unitary, dissipative processes.
Bassett \textit{et al.} applied the TDQT technique to extract the spin-orbit and spin-spin parameters of the NV center's excited state Hamiltonian, and to study the role of decoherence due to spontaneous photon emission, spectral diffusion, phonon-mediated orbital relaxation, hyperfine-induced spin dephasing, and the state-selective ISC transitions.

As a quantum control technique, the pulse timings can be chosen to achieve a desired unitary quantum operation on the ground-state spin.
If we are interested in the evolution within a qubit subspace (and assuming we can effectively isolate the evolution to that subspace within the excited state), we can view the effect of a pair of such pulses as a temporary change in the effective magnetic field.
With appropriate control over the pulse timings and excited-state Hamiltonian, this all-optical, and microwave-free technique can be applied to generate rotations for the ground-state spin qubit.

Consider for example the situation of the double-$\Lambda$ configuration of \cref{fig:CPT}(b) that is formed near an excited-state anticrossing of the $\ket{L,0}$ and $\ket{L,+1}$ eigenstates.
By tuning the polarization of the optical pulses following \cref{eq:CoupledES} such that the optically-coupled excited state is $\ket{E}=\ket{L}$, we can isolate most of the unitary dynamics to the four-dimensional subspace spanned by $\{\ket{G,0},\ket{G,+1}\}$ and $\{\ket{L,0},\ket{L,+1}\}$.
To model this, we start from a diagonal ground-state Hamiltonian
\begin{equation}
H_{\rm gs}  = \frac{\omega_{\rm gs}}{2} s_z,
\label{eq:Hgs}
\end{equation}
describing precession of the effective spin-1/2 qubit about the $z$ axis due to the effective external magnetic field with frequency $\omega_{\rm gs}$.
Here, $s_z$ is a spin-1/2 Pauli-$z$ operator acting on the $\{\ket{G,0},\ket{G,+1}\}$ spin subspace.
Similarly, the effective excited-state qubit Hamiltonian describes a precession about an axis tilted by an angle $\eta$ relative to the ground state, and with a different frequency $\omega_{\rm es}$,
\begin{equation}
H_{\rm es}  = \frac{\omega_{\rm es}}{2}\left(\sin\eta\,s_x+\cos\eta\,s_z\right)
 = \frac{\omega_{\rm es}}{2} s_z'.
\end{equation}
Here we have set the complex phase of the off-diagonal matrix element to zero, since in experiments this phase is convolved with the constant but unspecified relative timing between the optical pulses and the microwaves used to address the ground-state spin.

The full four-dimensional Hamiltonian of this effective model is
\begin{equation}
H  = \left(
\begin{array}{cc}
H_{\rm gs} & 0\\
0 & H_{\rm es}+\omega_{\rm opt}I
\end{array}\right)
= \frac{1}{2}\left(1-\sigma_z\right) H_{\rm gs} +\frac{1}{2}\left(1+\sigma_z\right)(H_{\rm es}+\omega_{\rm opt}I),
\label{eq:Hsimple}
\end{equation}
where $\omega_{\rm opt}$ is the optical frequency difference between $\ket{G}$ and $\ket{L}$, and $\sigma_z$ is a Pauli operator for the orbital GS-ES degree of freedom, i.e., $\sigma_z|GS\rangle=-|GS\rangle$ and
$\sigma_z|ES\rangle=|ES\rangle$.
The action of a resonant ultrafast pulse with polarization $\hat{H}$ (see \cref{fig:UFcontrol}) is described by the unitary operator of \cref{eq:PulseOperators}, which reduces to
\begin{equation}
  U_\mathrm{FP}(\theta,\phi) = \cos\left(\frac{\theta}{2}\right) - i\sin\left(\frac{\theta}{2}\right)(\cos(\phi)\sigma_x+\sin(\phi)\sigma_y),
\end{equation}
corresponding to a coherent rotation in the $\{\ket{G},\ket{L}\}$ orbital basis by an angle $\theta$ about an axis defined by $\ket{G}+e^{-i\phi}\ket{L}$ (i.e., an equatorial axis in the orbital Bloch sphere).

The excited-state Hamiltonian parameters $\eta$ and $\omega_\mathrm{es}$ can be tuned by the external magnetic, electric, and strain fields.
The effective expression, \cref{eq:HL}, for the $\ket{L}$-branch Hamiltonian is useful for identifying regimes in which unwanted mixing with other spin and orbital states are minimized.
\Cref{fig:UFcontrol}(b) shows the fine structure of $\ket{L}$ as a function of $B_z$ corresponding to the strain configuration ($\delta/h=\unit{6.7}{\giga\hertz}$, $\alpha_s=\unit{-0.08}{\radian}$) from Ref.~\cite{Bassett2014}.
The configuration is similar to the one we considered in \cref{sec:CPT-SRT}, where an avoided level crossing occurs between $\ket{L,0}$ and $\ket{L,+1}$ around $B_z=110$~G.
The existence of such an anticrossing is confirmed using standard photoluminescence excitation spectroscopy as in \cref{fig:UFcontrol}(c).
However, frequency-domain spectroscopy only provides information about the energy eigenvalues, not the eigenstates.

According to the Hamiltonian, the excited-state spin eigenstates are fully hybridized at the center of the anticrossing; hence the effective precession axis in the excited state is orthogonal to that of the ground state, lying in the equatorial plane of the qubit Bloch sphere.
At other field values, the precession axis is tilted by an angle $\eta$ that approaches zero far from the level anticrossing.
These eigenstates are directly revealed by TDQT measurements of the spin evolution between two ultrafast optical pulses, as shown in \cref{fig:UFcontrol}(d).
The figures show trajectories that begin from an initialized state in either $\ket{0}$ or $\ket{+1}$ (and, at $B=400$~G, from a spin superposition state).
The trajectories are fits to the raw TDQT data using an analytical model that captures both unitary and dissipative dynamics \cite{Bassett2014}.

At the center of the anticrossing where $\eta=\pi/2$, a full $\pi$-pulse on the spin qubit can be achieved using a single pair of optical pulses.
For sequences of multiple single-qubit operations, the relative phase between pulses is deterministically set by the pulse timings.
In this way, universal quantum operations on the spin can be achieved using pairs of identical optical pulses.
Furthermore, whereas the dispersive SRT technique is limited in this configuration to a Rabi frequency $\Omega_g\lesssim\unit{10}{\mega\hertz}\ll\Delta_2$, direct evolution in the excited state occurs at the bare coupling rate, $\Omega_g\sim\omega_es\sim\Delta_2$.
In the data of \cref{fig:UFcontrol}(d), $\Omega_g=\unit{260}{\mega\hertz}$, corresponding to a $\pi$ rotation in only \unit{1.9}{\nano\second}, which approaches the fastest operation times demonstrated for NV centers using microwaves \cite{Fuchs2009}.

\section{Conclusions and future directions\label{sec:Summary}}

The purpose of this chapter is to provide an introduction to quantum optics in the context of solid-state spins like the diamond NV center.
However, the methods and techniques we describe only scratch the surface of quantum optics and its potential applications for quantum information science.
For example, the CPT and SRT techniques described in \cref{sec:CPT-SRT} have been applied to realize alternate forms of robust quantum control employing geometric phases \cite{Yale2016,Zhou2017,Zhou2017a}.
Whereas we focused on the diamond NV center, the techniques are general and are now routinely applied to other quantum systems including quantum dots \cite{Greve2013} and other defect systems \cite{Rogers2014a,Pingault2014,Becker2016,Koehl2017,Becker2018,Christle2017}.
As the number of materials platforms and applications for spin-based quantum technologies expands \cite{Atature2018,Awschalom2018}, the importance of these techniques will only continue to grow.

\acknowledgments
The original work described in Refs.~\cite{Buckley2010,Yale2013,Bassett2014} was supported by the Air Force Office of Scientific Research, the Army Research Office, and the Defense Advanced Research Projects Agency.  Preparation of these notes and the associated lectures was supported by the U. S. National Science Foundation under CAREER award ECCS-1553511.  The author thanks David Hopper and S. Alex Breitweiser for their critical reading of this manuscript; Mario Agio, Irene D'Amico, Rashid Zia, and Costanza Toninelli for organizing the Enrico Fermi Summer School on \textit{Nanoscale Quantum Optics}; the Italian Physical Society for hosting the course at the Villa Monastero in Varenna; and all the students and lecturers who attended the school for many enjoyable discussions.

\bibliography{D:/References/NVdatabase,D:/References/SolidStateQIP}

\end{document}